\documentclass[11pt]{article}
\usepackage{amsmath, amssymb, graphicx, dsfont, xcolor, natbib, atbegshi}
\usepackage[margin=1in]{geometry}

\definecolor{darkgreen}{rgb}{0.0, 0.2, 0.13}
\definecolor{ao}{rgb}{0.0, 0.5, 0.0}
\definecolor{blush}{rgb}{0.87, 0.36, 0.51}

\DeclareMathOperator*{\expit}{expit}

\newtheorem{theorem}{Theorem}

\newtheorem{lemma}{Lemma}

\begin{document}

\title{An Approach to Nonparametric Inference on the Causal Dose Response Function}
\author{Aaron Hudson$^{1}$, Elvin H. Geng$^2$, Thomas A. Odeny$^3$, Elizabeth A. Bukusi$^3$,\\ Maya L. Petersen$^4$, Mark J. van der Laan$^4$
\\
\\
\small $^1$Vaccine and Infectious Disease Division, Fred Hutchinson Cancer Center
\\
\small$^2$Division of Infectious Diseases, Department of Medicine, Washington University in St. Louis
\\
\small$^3$Research Care Training Program, Centre for Microbiology Research, Kenya Medical Research Institute
\\
\small$^4$Division of Biostatistics, School of Public Health, University of California, Berkeley}
\date{}
\maketitle
\bibliographystyle{biom}

\begin{abstract}
The causal dose response curve is commonly selected as the statistical parameter of interest in studies where the goal is to understand the effect of a continuous exposure on an outcome.Most of the available methodology for statistical inference on the dose-response function in the continuous exposure setting requires strong parametric assumptions on the probability distribution. Such parametric assumptions are typically untenable in practice and lead to invalid inference. It is often preferable to instead use nonparametric methods for inference, which only make mild assumptions about the data-generating mechanism. We propose a nonparametric test of the null hypothesis that the dose-response function is equal to a constant function. We argue that when the null hypothesis holds, the dose-response function has zero variance. Thus, one can test the null hypothesis by assessing whether there is sufficient evidence to claim that the variance is positive. We construct a novel estimator for the variance of the dose-response function, for which we can fully characterize the null limiting distribution and thus perform well-calibrated tests of the null hypothesis. We also present an approach for constructing simultaneous confidence bands for the dose-response function by inverting our proposed hypothesis test. We assess the validity of our proposal in a simulation study. In a data example, we study, in a population of patients who have initiated treatment for HIV, how the distance required to travel to an HIV clinic affects retention in care.
\end{abstract}

\section{Introduction}

In many scientific studies, one of the main objectives is to use observational data to make inferences about the causal relationship between a treatment or exposure variable and some outcome.
Many commonly-used methods for causal inference make strong, and often untenable, parametric assumptions about the data's probability distribution.
Consequently, using such methods in practice can result in invalid inference under model mis-specification.
Interest has grown in instead using more robust nonparametric and semiparametric approaches methods for causal inference, which only make mild assumptions about the data-generating mechanism.

Our objective is to make nonparametric inference about the probability distribution of the counterfactual outcomes, or the collection of potential outcomes that would have been observed if a subject had received any possible level of the exposure \citep{rubin1974estimating}.
In many conventional analyses, it is of primary interest to make comparisons about the mean of the counterfactual outcome for different levels of the exposure.
The target estimand is the function that maps any given exposure level to the corresponding mean counterfactual outcome.
We commonly  refer to this estimand as the \textit{causal dose response function}.

The literature on nonparametric inference on mean counterfactual outcomes most commonly focuses on the setting in which the exposure is binary.
In this case, the mean counterfactual outcomes can, under standard causal assumptions, be characterized as pathwise differentiable estimands, or smooth functionals of the unknown probability distribution.
For estimands that have this characterization, there exist many strategies for constructing nonparametric estimators that converge to the true quantity at the parametric rate and achieve a tractable sampling distribution.
Commonly-used estimators for the mean counterfactual outcomes with binary exposures that satisfy these properties include the augmented inverse probability weighted estimator \citep{robins1995semiparametric} and the targeted maximum likelihood estimator \citep{van2006targeted}.
When such estimators are available, it is straightforward to perform statistical inference on the mean counterfactual outcomes by constructing confidence sets or performing hypothesis tests.
With minor modifications, these approaches for inference can be readily extended to the more general setting in which the exposure is a discrete random variable.

Nonparametric inference on the causal dose response function is more challenging when the exposure is a continuous random variable.
In this setting, the mean counterfactual outcomes are non-smooth functions of the probability distribution, and consequently, nonparametric estimation is not possible at the parametric rate.
There nonetheless exist proposals for consistent and rate optimal estimation of the dose response function in a nonparametric model.
For instance, \cite{diaz2013targeted} introduce a cross-validated targeted minimum loss-based estimator for the dose response function, \cite{kennedy2017non} and \cite{colangelo2020double} propose nonparametric kernel smoothing estimators, and \cite{westling2020causal} proposes an estimator based on isotonic regression.
Rate optimal nonparametric estimators, such as those described above, typically retain a non-negligible asymptotic bias.
Due to this bias retention, these estimators attain non-standard limiting distributions, and constructing hypothesis tests or confidence sets based on the estimators can be challenging.
In order to obtain an estimator for the dose response curve that converges at the parametric rate, it is necessary to make parametric assumptions about either the conditional distribution of the exposure given the covariates \citep{hirano2004propensity, imai2004causal, galvao2015uniformly} or the conditional mean of the outcome given exposure and covariates \citep{robins2000marginal, zhang2016causal}.
These approaches may not be preferred because specifying a parametric model correctly can be challenging in practice.

There exist alternative approaches to nonparametric causal inference with continuous exposures which are based on the study of a finite-dimensional parameter that summarizes the effect of the exposure on the outcome \citep{neugebauer2007nonparametric, munoz2012population, hines2021parameterising}.
While such summaries can be informative and carry meaningful interpretations, they may not suffice in settings where an investigator is primarily interested in learning about the dose response function in its entirety.

Our work draws upon some recent advances in nonparametric inference on the dose response function and inference on non-pathwise differentiable estimands more broadly.
\cite{westling2021nonparametric} demonstrates that one can construct an omnibus nonparametric test of the hypothesis that the dose response function is flat by simultaneously estimating primitive functions, or integrals, of the dose response function.
\cite{hudson2021inference} uses a generalization of this idea to develop a framework for testing hypotheses about non-pathwise differentiable estimands by estimating a suitably large collection of pathwise differentiable estimands that can effectively summarize the target.
\cite{hudson2021inference} also proposes a method for constructing simultaneous confidence sets for the target estimand by inverting the proposed hypothesis test.

In this paper, we propose a novel method for inference on the dose response function using the framework of \cite{hudson2021inference}.  
We develop a test of the null hypothesis that a mean-centered version of the dose response function is equal to any given null function.
Of particular importance is the instance where the null function is zero, in which case we perform a test of the null hypothesis that the dose response function is flat.
We also discuss construction of simultaneous confidence sets for the centered dose response function, and we describe a method for summarizing the confidence sets.

The remainder of the paper is organized as follows. 
In Section 2, we provide a high-level overview of our proposed methodology, and in Section 3, we discuss our inferential procedure in more detail and describe its theoretical properties.
In Section 4, we discuss implementation of our method.
In Section 5, we study the behavior of our proposal in a simulation study.
In Section 6, we apply our method to data from the Adaptive Strategies for Preventing
and Treating Lapses of Retention in HIV Care (ADAPT-R) trial (NCT02338739), which studies retention in care for people who initiated treatment for HIV.
We conclude with a brief discussion in Section 7.

\section{Overview}

\subsection{Identification of the Dose Response Function} \label{overview-identify}

Let $O_1, \ldots, O_n$ be \textit{i.i.d.} random vectors drawn from a probability distribution $P_0$, which resides in a model $\mathcal{M}$. We allow $\mathcal{M}$ to be a rich nonparametric model that is essentially unrestricted and must only satisfy some mild regularity conditions.
We express our data as $O = (W, A, Y)$, where $Y$ is a bounded real-valued outcome with sample space  $\mathcal{Y} \subset \mathbb{R}$, $A$ is a bounded exposure variable with sample space $\mathcal{A} \subset \mathbb{R}$, and $W$ is a $q$-dimensional vector of covariates, with sample space $\mathcal{W} \subset \mathbb{R}^q$.
Throughout, we write $E_P$ to denote the expectation under any probability distribution $P \in \mathcal{M}$, and we use the shorthand notation $E_0 := E_{P_0}$.

Let $Y(a)$ denote the counterfactual outcome under exposure level $a$, or the potential outcome that would have been observed if an individual had been observed with exposure level $A = a$.
We define the counterfactual mean outcome as the mapping $P \mapsto \theta_P(a) := E_P[Y(a)]$ for $a \in \mathcal{A}$.
Our objective is to perform inference on $\theta_0 := \theta_{P_0}$, which is commonly called the \textit{causal dose response function}.

The stochastic process of counterfactual outcomes for subject $i$, $\{Y_i(a): a \in \mathcal{A}\}$ is not observable, as we only observe the outcome under one single exposure level $A_i$.
However, under some standard causal assumptions, the counterfactual mean outcomes can nonetheless be estimated from the data.
Let $Q_{P}(w,a) := E_P[Y|W = w,A=a]$ denote the conditional mean of $Y$ given the exposure and covariates, and let $g_P(a|w) := \frac{d}{da} P(A \leq a|W=w)$ be the conditional density of the exposure given the covariates.
Similarly as above, we use the shorthand notation $Q_0 := Q_{P_0}$ and $g_{0} := g_{P_0}$.
The dose response function can be estimated from the data under the following assumptions:
\begin{itemize}
\item[] \textbf{Assumption A1 (Consistency)}. $A = a$ implies $Y = Y(a)$.
\item[] \textbf{Assumption A2 (No unemasured confoundedness).} $E_0[Y(a)|W = w] = Q_0(w,a)$ for all $a \in \mathcal{A}$, $w \in \mathcal{W}$.
\item[] \textbf{Assumption A3 (Positivity).} There exists $g_{\min} > 0$ so that $\inf_{a \in \mathcal{A}} g_0(a|w) > g_{\min}$ for all $w \in \mathcal{W}$.
\end{itemize}

The consistency assumption states that the observed outcome for each subject is the potential outcome under their observed exposure level.
The no unmeasured confoundedness assumption says that all confounding variables are contained within $W$.
The positivity assumption means that it is possible for subjects with any covariate measurement to receive any treatment $a \in \mathcal{A}$.
When Assumptions A1-A3 hold, the counterfactual mean outcome at any exposure level $a$ can be expressed as
\begin{align*}
\theta_0(a) = E_{0,W}[Q(W,a)],
\end{align*}
where $E_{0,W}$ denotes the expectation with respect to the marginal distribution of $W$ under $P_0$.

Assumptions A1-A3 typically hold in randomized trials, but they are generally unverifiable in observational studies.
However, even when the causal assumptions fail, $\theta_0$ retains an interpretation as a mean regression function, which can be used to study the conditional association between the exposure and the outcome, given the covariates.

\subsection{Summary of Proposed Methodology} \label{overview-summary}

We first derive a test of the null hypothesis that the dose response function, centered about its mean, is equal to a given candidate null function.
Let $\bar{\theta}_P: a \mapsto \theta_P(a) - E_P[\theta_P(A)]$ be the mean-centered dose response function.
For an arbitrary function $\theta^*$ from $\mathcal{A}$ to $\mathbb{R}$, let $\bar{\theta}^*_{P} = \theta^* - E_P[\theta^*(A)]$, and let $\bar{\theta}^*_0 := \bar{\theta}^*_{P_0}$.
We are interested in testing the null hypothesis that $\bar{\theta}_0$ is equal to the candidate null parameter $\bar{\theta}^*_0$,
\begin{align}
H_0: \bar{\theta}_0(a) = \bar{\theta}^*_0(a), \text{ for all } a \in \mathcal{A}.
\label{flatnull}
\end{align}
Of particular interest in many applications is the case where $\bar{\theta}_* = 0$, which corresponds to the null hypothesis that the dose response function is flat, or that the counterfactual mean outcomes are the same at every level of the exposure.

We also propose an approach for constructing confidence sets for the centered dose response function.
That is, for any $\alpha \in (0,1)$, we construct a set $\mathcal{C}^\alpha_n$ that contains $\bar{\theta}_0$ with probably at least $1-\alpha$ as $n$ tends to infinity, i.e.,
\begin{align*}
\liminf_{n \to \infty}P_0\left(\bar{\theta}_0 \in \mathcal{C}^\alpha_n\right) \geq 1 - \alpha.
\end{align*}

When it is possible to obtain an estimator for $\theta_0$ that has negligible asymptotic bias and a characterizable limitng distribution, it is straightforward to derive hypothesis tests and construct confidence sets based on the estimator.
When the dose response function is pathwise differentiable, meaning that $\theta_P(a)$ changes smoothly in $P$ with respect to local fluctuations around $P_0$, there is well-established theory and methodology for constructing $n^{1/2}$-consistent and asymptotically efficient estimators \citep{bickel1998efficient}.
However, $\theta_P(a)$ is typically only pathwise differentiable in a nonparametric model when the event $\{A = a\}$ occurs with probability greater than zero.
While this condition can be satisfied when $A$ has a discrete support under $P_0$, it is generally not met when the exposure variable is continuous.

While the dose response function is not pathwise differentiable when the exposure is continuous, one can perform inference by instead estimating a collection of pathwise differentiable functionals of the dose response function, as suggested by, e.g., \cite{hudson2021inference} and \cite{westling2021nonparametric}.
In this work, we present an approach based on estimation of linear transformations of the causal dose response function.
In what follows, we briefly summarize our approach to inference.

For any function $h$ from $\mathcal{A}$ to $\mathbb{R}$, we define 
\begin{align*}
\psi_{P, \theta^*}(h) := E_P\left[ \left\{\bar{\theta}_P(A) - \bar{\theta}^*_P(A)\right\} h(A) \right]
\end{align*}
as the $L_2(P)$ inner product of $\bar{\theta}_P - \bar{\theta}^*_P$ and $h$, and let $\psi_{0, \theta^*}(h) := \psi_{P_0, \theta^*}(h)$.
Observe that if $\bar{\theta}_0$ is equal to $\bar{\theta}^*_0$ almost everywhere,  $\psi_{0, \theta^*}(h) = 0$ for all functions $h$, and if $\bar{\theta}_0$ is not almost everywhere equal to $\bar{\theta}^*_0$, there exists a function $h_*$ such that $\psi_{0, \theta^*}(h_*) \neq 0$.
Therefore, one could test the null hypothesis in \eqref{flatnull} by assessing whether, for a large class $\mathcal{H}$ of bounded functions from $\mathcal{A}$ to $\mathbb{R}$,
\begin{align}
\Psi_{0, \theta^*}(\mathcal{H}) := \sup_{h \in \mathcal{H}}\left|\psi_{0, \theta^*}(h)  \right| = 0,
\label{sup-lin-combo}
\end{align}
or in other words, whether there exists a linear transformation of $\bar{\theta}_0 - \bar{\theta}^*_0$ that is non-zero.

We develop a test of the null hypothesis based on estimation of $\Psi_{0,\theta^*}(\mathcal{H})$.  We later show that $\psi_{0,\theta^*}(h)$ is pathwise differentiable for any fixed $h$ and can therefore be estimated at the parametric rate of $n^{1/2}$.  If, in addition, $\mathcal{H}$ is not too complex, one can construct an estimator $\psi_{n, \theta^*}$ such that the standardized process $\left\{n^{1/2}\left[\psi_{n, \theta^*}(h) - \psi_{0, \theta^*}(h)  \right]: h \in \mathcal{H} \right\}$ converges weakly to a Gaussian process $\mathbb{G} := \left\{\mathbb{G}(h): h \in \mathcal{H}\right\}$.
As a consequence, we have that $\Psi_{n, \theta^*}(\mathcal{H}) := \sup_{h \in \mathcal{H}} \left|\psi_{n, \theta^*}(h)\right|$ is a consistent estimator for $\Psi_{0, \theta^*}(\mathcal{H})$, and under the null, when $\psi_{0,\theta^*}(h) = 0$ for all $h$, $n^{1/2}\Psi_{n, \theta^*}(\mathcal{H})$ converges weakly to $\sup_{h \in \mathcal{H}}\left|\mathbb{G}(h)\right|$.
Because $\Psi_{n,\theta^*}(\mathcal{H})$ has a tractable limiting distribution, it can be used as a test statistic.
A hypothesis test that rejects the null when $n^{1/2}\Psi_{n, \theta^*}(\mathcal{H})$ is larger than the $1-\alpha$ quantile of the null limiting distribution would achieve the type-1 error level $\alpha$ asymptotically.

We can invert our proposed hypothesis test to obtain a confidence set for $\bar{\theta}_0$.
Let $\Theta$ be a large nonparametric class of functions from $\mathcal{A}$ to $\mathbb{R}$ that serve as candidate values for the dose response function.
Consider the set
\begin{align}
\mathcal{C}^\alpha_n := \left\{\bar{\theta}^*_0: \theta^* \in \Theta, \text{ and we fail to reject } \bar{\theta}_0 = \bar{\theta}^*_0 \text{ at the level } \alpha \text{ based on } O_1, \ldots, O_n \right\}.
\label{conf-set}
\end{align}
We can interpret $\mathcal{C}^\alpha_n$ as the set of functions in $\Theta$ that are accordant with the observed data.
If $\theta_0$ belongs to $\Theta$, then $\bar{\theta}_0$ belongs to $\mathcal{C}^{\alpha}_n$ with probability tending $1-\alpha$, and $\mathcal{C}^{\alpha}_n$ is an asymptotically valid confidence set for $\bar{\theta}_0$.
Even when $\theta_0$ does not belong to $\Theta$, $\mathcal{C}^\alpha_n$ retains a nice interpretation as long as $\Theta$ contains a good approximation for $\theta_0$.

The confidence set $\mathcal{C}_n^\alpha$ can be difficult to visualize as it is a complex set of infinite-dimensional objects.
We propose to simply summarize $\mathcal{C}_n^\alpha$ by displaying the smallest band that contains all functions in $\mathcal{C}_n^\alpha$, which can be defined point-wise as
\begin{align}
C_n^\alpha(a) :=
\left[\inf \left\{\bar{\theta}^*_0(a): \bar{\theta^*_0} \in \mathcal{C}_n^\alpha\right\}, \sup \left\{\bar{\theta}^*_0(a): \bar{\theta^*_0} \in \mathcal{C}_n^\alpha\right\}\right].
\label{conf-band}
\end{align}

The inferential procedure described above requires selection of the class of functions $\mathcal{H}$, 
the elements of which index the set of linear transformations of $\bar{\theta}_0 - \bar{\theta}^*_0$ that we need to estimate.
Asymptotic type-1 error control is preserved using any choice $\mathcal{H}$ that satisfies some mild regularity conditions, to be discussed later.
The choice of $\mathcal{H}$ does, however, influence the test's statistical power.
We use the following intuition for constructing an $\mathcal{H}$ so that our test is well-powered.
Because the inner product $\psi_{0, \theta^*}(h)$ is a measure of orthogonality of $\bar{\theta}_0 - \bar{\theta}^*_0$ to a given $h$, the amount of evidence $\psi_{0, \theta^*}(h)$ provides against the null depends on the \textit{shape} of $h$, rather than the \textit{scale}.
It is therefore sensible to construct $\mathcal{H}$ as a class of equally-scaled functions that contains the function that is least orthogonal $\bar{\theta}_0$.
We scale each $h$ to have unit variance; while other measures of scale could be chosen, we scale by the standard deviation for simplicity.
It can be shown by an application of the Cauchy-Schwarz inequality that the maximizer of $|\psi_{0, \theta^*}(h)|$, among all $h$ with unit standard deviation, can be expressed as
\begin{align}
h_0 := \text{Var}_0(\theta_0(A) - \theta^*(A))^{-1/2} \left(\bar{\theta}_0 - \bar{\theta}^*_0\right),
\label{steepest-h}
\end{align}
where $\text{Var}_0$ is the population variance under $P_0$.
The maximizer $h_0$ is a scaled difference between the true dose response function and the candidate null parameter, and the maximal inner product $\psi_{0,\theta^*}(h_0)$ is equal to the standard deviation of $\theta_0(A) - \theta^*_0(A)$.
We construct $\mathcal{H}$ as a class of smooth functions that contains a good approximation of $h_0$ so that our target estimand $\Psi_{0, \theta^*}(\mathcal{H})$ can be interpreted as an approximation of the standard deviation of the difference between the truth and the null.


The hypothesis testing procedure proposed by \cite{westling2021nonparametric} can be viewed as a special case of our method with $\mathcal{H}$ taken as the class of binary indicators of whether an input exposure level $a$ is greater than any given cutoff $a_0 \in \mathcal{A}$.
While \cite{westling2021nonparametric} shows that this choice results in an omnibus test of the null hypothesis, we later show that in finite samples, performance can be improved by considering a class of functions that contains the function that is least orthogonal to the difference between the true dose response function and the null.

Our proposal is closely related to the nonparametric score test presented in \cite{hudson2021inference}.
The authors use the representation of function-valued non-pathwise differentiable parameters as the minimizer of a population risk functional to derive a set of estimating equations, indexed by functions $h$ in a large class $\mathcal{H}$, that the true population parameter must satisfy.
They then construct hypothesis tests that assesses whether the candidate null parameter also satisfies these estimating equations.
Their proposal rejects the null hypothesis when the data provides contrary evidence.
Our proposal can be viewed as a special case of this approach, where we assess whether $\theta^*$ satisfies the estimating equations $\psi_{0, \theta^*}(h) = 0$ for all $h \in \mathcal{H}$.

\section{Inferential Procedure}

Having provided a brief overview of our inferential procedure in the previous section, we now provide theoretical details.
In this section, we first discuss estimation of $\psi_{0,\theta^*}(h)$, and we subsequently describe an approach for determining whether there is sufficient evidence to conclude whether $\Psi_{0, \theta^*}(\mathcal{H})$ is zero.

\subsection{Estimation of $\psi_{0,\theta^*}(h)$} \label{methods-estimator}

Recall that our proposal requires us to have available an estimator $\psi_{n}(h)$ of $\psi_{0}(h)$ and a class of functions $\mathcal{H}$ such that the process $\left\{n^{1/2}\left[\psi_{n}(h) - \psi_{0}(h)\right]: h \in \mathcal{H}\right\}$ converges weakly to a Gaussian process.
In this subsection, we specify conditions on $\psi_{n, \theta^*}$ and $\mathcal{H}$ such that the weak convergence property is satisfied, and we discuss construction of a weakly convergent estimator.

As noted in Section 2, for any estimand that is pathwise differentiable in the sense of \cite{bickel1998efficient}, one can construct an estimator that, when centered around the true target estimand, converges weakly to a Gaussian distribution at the parametric rate of $n^{1/2}$.
Constructing such an estimator and establishing efficiency typically requires knowledge of the \textit{efficient influence function} of the estimand of interest.
The following lemma states that $\psi_{P, \theta^*}(h)$ is pathwise differentiable in a nonparametric model and provides the form of the efficient influence function.

\begin{lemma}
The parameter $\psi_{P, \theta^*}(h)$ is pathwise differentiable in a nonparametric model, and its nonparametric efficient influence function is 
\begin{align*}
\phi_{P, \theta^*}(w,a,y; h) := &\left[\bar{\theta}_P(a) - \bar{\theta}^*_P(a)  + \frac{E_P[ g_P(a|W)]}{g_P(a|w)} \left\{y - Q_P(w,a)   \right\} \right]\left\{h(a) - E_P[h(A)]\right\}
 +
\\
&E_P\left[Q_P(w,A)\left\{h(A) - E_P[h(A)]\right\}\right] - \left\{2E_P[\bar{\theta}_P(A)h(A)] - E_P[\bar{\theta}^*_P(A)h(A)]\right\}.
\end{align*}
\end{lemma}
As before, we use the shorthand notation $\phi_{0,\theta^*} := \phi_{P_0, \theta^*}$ to denote the efficient influence function at $P_0$.
Lemma 1 generalizes a result presented in \cite{westling2021nonparametric} that characterizes the efficient influence function for the special case in which $h$ is an indicator of whether the observed treatment level is greater than a specified cutoff, and $\bar{\theta}^*_0 = 0$.

Because $\psi_{0,\theta^*}(h)$ is pathwise differentiable, it possible to construct an estimator $\psi_{n, \theta^*}(h)$ that is asymptotically linear in the sense that
\begin{align}
\psi_{n, \theta^*}(h) - \psi_{0,\theta^*}(h) = \frac{1}{n}\sum_{i=1}^n \phi_{P_0,\theta^*}(O_i; h) + r_n(h),
\label{asymplin}
\end{align}
where $r_n(h) = o_P(n^{-1/2})$.
In view of the central limit theorem and the fact that $\phi_{0,\theta^*}$ has zero mean and finite variance, an asymptotically linear estimator $\psi_{n, \theta^*}(h)$ is asymptotically Gaussian for any fixed $h$.
However, our proposal requires a stronger notion of uniform convergence of $\psi_{n,\theta^*}(h)$ for $h$ in a large collection $\mathcal{H}$, so some additional conditions are needed.
The following lemma, which is a consequence of Slutsky's theorem, provides conditions under which the desired uniform convergence holds.
\begin{lemma}
Let $\psi_{n,\theta^*}(h)$ be an asymptotically linear estimator of $\psi_0(h)$ that has the representation in \eqref{asymplin} for any $h \in \mathcal{H}$, and let $\ell^\infty(\mathcal{H})$ denote the vector space of bounded real-valued functionals on $\mathcal{H}$.
Assume the following conditions hold:
\begin{enumerate}
\item $\left\{\phi_{0, \theta^*}(\cdot ;h): h \in \mathcal{H}\right\}$ is a $P_0$-Donsker class,
\item $\sup_{h \in \mathcal{H}}|r_n(h)| = o_P(n^{-1/2})$.
\end{enumerate}
Then $\left\{n^{1/2}\left[\psi_{n, \theta^*}(h) - \psi_{0, \theta^*}(h)\right]: h \in \mathcal{H}\right\}$ converges weakly to a tight Gaussian process $\mathbb{G}$ as an element of $\ell^\infty(\mathcal{H})$, where $\mathbb{G}$ has mean zero and covariance $\Sigma:(h_1, h_2) \mapsto E_0[\phi_{0, \theta^*}(O;h_1)\phi_{0, \theta^*}(O; h_2)]$.
\end{lemma}
The first condition of Lemma 2 is a constraint on the complexity of $\mathcal{H}$ and typically holds when $\mathcal{H}$ is a $P_0$-Donsker class.
The second condition directly involves the estimator $\psi_n$ and requires that the remainder term is asymptotically negligible in a uniform sense.

In what follows, we present two strategies for constructing weakly convergent estimators for $\{\psi_{0,\theta^*}(h): h \in \mathcal{H}\}$. 
We begin by considering a na\"ive plug-in estimator of $\psi_0(h)$.
Suppose we have available a consistent estimator $\hat{P}_n$ for $P_0$.
In practice, we do not need to estimate the entire distribution $P_0$, and we must only estimate nuisance parameters upon which $\psi_{0,\theta^*}$ and $\phi_{0, \theta_*}$  depend.
In our setting, there are four nuisance parameters: (i) $F_{0, W}$, the marginal cumulative distribution for $W$, (ii) $F_{0, A}$ the marginal cumulative distribution function for $A$, (iii) $Q_{0}$ the conditional mean of $Y$ given $A$ and $W$, and (iv) $g_0$ the conditional density of $A$ given $W$.
One can obtain estimators $F_{n, W}$  and $F_{n, A}$ for $F_{0,W}$ and $F_{0, A}$  nonparametrically using the empirical distribution function, and one typically requires machine learning to construct consistent nonparametric estimators $Q_{n}$ and $g_n$ for $Q_0$ and $g_0$.
Given the initial estimator $\hat{P}_n$, one can obtain the na\"ive plug-in estimator $\psi_{\hat{P}_n}(h)$, which can be expressed as
\begin{align*}
\psi_{\hat{P}_n, \theta^*}(h) = n^{-1} \sum_{i=1}^n \left[ \theta_n(A_i) - \theta^*(A_i) - n^{-1}\sum_{j=1}^n\left\{\theta_n(A_j) - \theta^*(A_j)\right\}  \right] h(A_i),
\end{align*}
where $\theta_n: a \mapsto n^{-1}\sum_{i=1}^n Q_n(a, W_i)$ is the plug-in estimator for the dose response function.

The plug-in estimator $\psi_{\hat{P}_n, \theta^*}(h)$ will typically retain non-negligible asymptotic bias for $\psi_{0, \theta^*}(h)$ and consequently will not be asymptotically linear.
This bias is attributable to the fact that nonparametric estimators $Q_n$ of $Q_0$ are usually obtained by balancing a bias-variance tradeoff that is sub-optimal for the objective of estimating $\psi_{0, \theta^*}(h)$.
We discuss two widely-used strategies for correcting the bias of the na\"ive plug-in: one-step estimation \citep{pfanzagl1982contributions} and targeted minimum loss-based estimation \citep{van2011targeted, van2018targeted}.

The estimation strategies we discuss require the following assumptions:
\begin{enumerate}
\item[] \textbf{Assumption B1.} There exists a $P_0$-Donsker class $\Phi$ that contains $\phi_{0, \theta^*}(\cdot; h)$ and $\phi_{\hat{P}_n, \theta^*}(\cdot; h)$ for each $h \in \mathcal{H}$ with probability tending to one.
\item[] \textbf{Assumption B2.} The nuisance parameter estimators satisfy:
\begin{align*}
&\int \left\{Q_n(w,a) - Q_0(w,a)\right\}^2 dP_0(w,a) = o_P(1),
\\
&\int \left\{g_n(a|w) - g_0(a|w)\right\}^2 dP_0(w,a) = o_P(1),
\\
&\int \left|\left\{Q_n(w,a) - Q_0(w,a)\right\}\left\{g_n(a|w) - g_0(a|w)\right\}\right|dP_0(w,a) = o_P(n^{-1/2}).
\end{align*}
\end{enumerate}
Assumptions B1 and B2 impose conditions on the estimators for the nuisance parameters upon which the target estimand and the efficient influence function depend.  
Assumption B1 places a constraint on the complexity of the family of candidate estimators for the nuisance components.
Many flexible nonparametric estimators, e.g., those constructed via the highly adaptive LASSO \citep{benkeser2016highly}, satisfy this condition.
Assumption B2 places a requirement on the rates of convergence that the conditional mean and conditional density estimators must achieve.  
This condition holds when both estimators are consistent, and the product of the convergence rates is greater than $n^{1/2}$.
In view of Assumption B2, the estimators of $\psi_{0,\theta^*}(h)$ we develop are doubly robust in the sense that consistency and asymptotic normality are achieved if one of the nuisance parameters is estimated at a slow rate as long as the other nuisance is estimated at a fast enough rate to compensate.

We first construct a one-step estimator for $\psi_{0,\theta^*}(h)$.
Given an estimator for the nuisance parameters upon which $\phi_{0, \theta^*}(h)$ depends, we can obtain an estimator $\phi_{\hat{P}_n, \theta^*}(\cdot; h)$ for the efficient influence function $\phi_{0, \theta^*}(\cdot;h)$.
The empirical average of the estimator of the influence function $\phi_{\hat{P}_n, \theta^*}$ can be shown to serve as a first-order approximation to the bias of the na\"ive plug-in \citep{pfanzagl1982contributions}.
This allows for one to perform a, so-called, one-step bias correction.
We define the one-step estimator as
\begin{align}
\psi^{\mathrm{I}}_{n, \theta^*}(h) &= \psi_{\hat{P}_n, \theta^*}(h) + n^{-1}\sum_{i=1}^n \phi_{\hat{P}_n, \theta^*}(O_i; h) \nonumber
\\
&=   \psi_{\hat{P}_n, \theta^*}(h)+ n^{-1}\sum_{i=1}^n \left[\frac{n^{-1}\sum_{j=1}^n g_n(A_i|W_j)}{g_n(A_i|W_i)} \left\{Y_i - Q_n(W_i,A_i)   \right\} \right]\left\{h(A_i) - n^{-1}\sum_{j=1}^n h(A_j)\right\}.
\label{one-step}
\end{align}
The following theorem states that, under mild regularity conditions, the one-step estimator is asymptotically linear and hence asymptotically Gaussian.
\begin{theorem}
Under assumptions $B1$ and $B2$, the one-step estimator $\psi_{n,\theta^*}^{\mathrm{I}}(h)$ has the asymptotically linear representation in \eqref{asymplin}, with $\sup_{h \in \mathcal{H}}|r_n(h)| = o_P(n^{-1/2})$.
\end{theorem}

While one-step estimators are asymptotically efficient, they are not guaranteed to be compatible in the sense that there exists a probability distribution $P$ in the model $\mathcal{M}$ such that $\psi_{P, \theta^*}(h) = \psi_{n, \theta^*}^{\mathrm{I}}(h)$ for all $h$ in  $\mathcal{H}$. 
Targeted minimum loss-based (TML) estimation is an appealing alternative strategy that can be used to construct an estimator for $\{\psi_{0, \theta^*}(h): h \in \mathcal{H}\}$ that is compatible in this sense.
TML estimators correct for the bias of the na\"ive plug-in by updating the initial estimator $\hat{P}_n$ of $P_0$ to obtain a new estimator $\tilde{P}_n$ such that the updated plug-in $\left\{\psi_{\tilde{P}_n, \theta^*}: h \in \mathcal{H}\right\}$ that takes as input $\tilde{P}_n$ has reduced bias for the target estimand $\{\psi_{0, \theta^*}(h): h \in \mathcal{H}\}$. 
Therefore, as long as $\tilde{P}_n$ resides within $\mathcal{M}$, the TML estimator is compatible.
In our presentation, we only briefly summarize some of the main principles of targeted learning, and we refer readers to \cite{van2011targeted, van2018targeted} for a comprehensive discussion.

The main idea behind targeted minimum loss-based estimation is to construct an updated estimator $\tilde{P}_n$ based on the initial estimator $\hat{P}_n$ so that the following efficient influence function estimating equations are satisfied:
\begin{align*}
\sup_{h \in \mathcal{H}} \left|n^{-1} \sum_{i=1}^n \phi_{\tilde{P}_n, \theta^*}(O_i; h)\right| = o_P(n^{-1/2}),
\end{align*}
and such that $\tilde{P}_n$ remains sufficiently close to $\hat{P}_n$, so as to remain a good estimator for $P_0$.
Because estimating a marginal distribution function using the empirical distribution function does not generate bias for the target estimand, we do not need to update the initial estimators $F_{A,n}$ and $F_{W,n}$ for $F_{A,0}$ and $F_{W,0}$.
We only need to obtain an updated estimator $\tilde{Q}_n$ for the conditional mean $Q_0$, since the initial estimator makes a bias-variance trade-off that is suboptimal for estimation of the target parameter.

It can be verified algebraically that for any choice $\tilde{Q}_n$, the empirical average of the efficient influence function evaluated at $\tilde{P}_n$ can be expressed as
\begin{align*}
n^{-1} \sum_{i=1}^n \phi_{\tilde{P}_n, \theta^*}(O_i; h) = n^{-1} \sum_{i=1}^n \left\{Y_i - \tilde{Q}_n(W_i,A_i)\right\}Z_{n}(W_i, A_i;h),
\end{align*}
where we define $Z_{n}(w,a;h)$ as
\begin{align}
Z_{n}(w, a;h) := \frac{n^{-1}\sum_{i=1}^n g_n(a|W_i)}{g_n(a|w)}\left\{h(a) - n^{-1}\sum_{i=1}^n h(A_i)\right\}.
\label{Zn}
\end{align}
Thus, $\tilde{Q}_n$ satisfies the efficient influence function estimating equations at an adequate level if
\begin{align}
\sup_{h \in \mathcal{H}} \left| n^{-1} \sum_{i=1}^n \left\{Y_i - \tilde{Q}_n(W_i,A_i)\right\}Z_{n}(W_i, A_i; h) \right| = o_P(n^{-1/2}).
\label{eif-solve}
\end{align}

We now discuss how to construct a $\tilde{Q}_n$ that satisfies \eqref{eif-solve}.
Let $Q_{n, \beta}$ be a parametric working model indexed by a scalar parameter $\beta \in \mathbb{R}$, for which $Q_{n,\beta} = Q_n$ when $\beta = 0$.
We construct the working model so that the derivative of the squared error loss is equal in magnitude to the supremum over $\mathcal{H}$ of the empirical average of $\phi_{\tilde{P}_n}(\cdot; h)$ with $\tilde{Q}_n = Q_{n, \beta}$, i.e.,
\begin{align}
\left|\frac{d}{d\beta}\left[ (2n)^{-1}\sum_{i=1}^n\left\{Y_i - Q_{n,\beta}(W_i, A_i) \right\}^2\right]\right| =
\sup_{h \in \mathcal{H}} \left| n^{-1}\sum_{i=1}^n\left\{Y_i - Q_{n,\beta}(W_i, A_i) \right\}Z_{n}(W_i, A_i; h)\right|.
\label{unif-lfsm}
\end{align}
We then take $\tilde{Q}_n = Q_{n, \beta_n}$, where $\beta_n$ is a near minimizer of the squared error loss and satisfies
\begin{align*}
n^{-1} \sum_{i=1}^n \left\{Y_i - Q_{n,\beta_n}(W_i,A_i)\right\}^2 = \underset{\beta \in \mathbb{R}}{\inf} \,\,\left[ n^{-1} \sum_{i=1}^n \left\{Y_i - Q_{n,\beta}(W_i,A_i)\right\}^2\right] + \delta_n,
\end{align*}
for a small positive sequence $\delta_n \downarrow 0$.
Because $\tilde{\beta}_n$ is a near minimizer of the loss, we can see that this choice of $\tilde{Q}_{n}$ satisfies \eqref{eif-solve} for $\delta_n$ sufficiently small.
We note that a sub-model satisfying \eqref{unif-lfsm} is referred to as a \textit{universal least favorable submodel} and provides the maximal reduction of the bias of $\{\psi_{\tilde{P}_n,\theta^*}(h): h \in \mathcal{H}\}$ as $Q_{n,\beta}$ moves away from $Q_n$ toward $Q_{n, \beta_n}$.
Strategies based on a locally least favorable submodel, which would satisfy \eqref{unif-lfsm} only when $\beta = 0$, could alternatively have been considered, but such approaches tend to perform worse in small samples, in particular when the target estimand is multidimensional or infinite-dimensional \citep{van2016one}.
We recursively define the universal least favorable sub-model $Q_{n, \beta}$ point-wise as
\begin{align}
Q_{n, \beta}(w,a) = 
\begin{cases} 
      Q_n(w,a
      	) + \int_0^\beta Z_{n}(a,w;h_{n, b}) db & \beta \geq 0 \\
      Q_n(w,a) - \int_{\beta}^0 Z_{n}(w,a;h_{n, b}) db & \beta < 0
   \end{cases},
\label{ulfsm}
\end{align}
where $h_{n,\beta}$ is a solution to
\begin{align*}
&\left| n^{-1}\sum_{i=1}^n\left\{Y_i - Q_{n,\beta}(W_i, A_i)\right\}Z_{n}(W_i,A_i;h_{n,\beta})\right|
=
\sup_{h \in \mathcal{H}} \left| n^{-1}\sum_{i=1}^n\left\{Y_i - Q_{n,
\beta}(W_i,A_i)\right\}Z_{n}(W_i,A_i;h)\right|.
\end{align*}
It can be verified using the fundamental theorem of calculus that the above working model satisfies \eqref{unif-lfsm}.

After obtaining the updated estimator $\tilde{Q}_n$, we can construct the TML estimator as
\begin{align}
\psi_{n, \theta^*}^{\mathrm{II}}(h) = \psi_{\tilde{P}_n, \theta^*}(h) = n^{-1} \sum_{i=1}^n \left[ \theta_{\tilde{P}_n}(A_i) - \theta^*(A_i) - n^{-1}\sum_{j=1}^n\left\{\theta_{\tilde{P}_n}(A_j) - \theta^*(A_j)\right\}  \right] h(A_i),
\label{tmle}
\end{align}
where $\theta_{\tilde{P}_n}: a \mapsto n^{-1}\sum_{i=1}^n \tilde{Q}_n(W_i, a)$ is the updated TML estimator for the dose response function.
The following theorem states that the TML estimator satisfies the conditions of Lemma 2.
\begin{theorem}
Under assumptions B1 and B2, the targeted minimum loss-based estimator $\psi^{\mathrm{II}}_{n,\theta^*}(h)$ has the asymptotically linear representation in \eqref{asymplin} with $\sup_{h \in \mathcal{H}}|r_n(h)| = o_P(n^{-1/2})$.
\end{theorem}

\subsection{Inference on $\Psi_{0,\theta^*}(\mathcal{H})$} \label{methods-test}

We are at this point prepared to discuss inference on $\Psi_{0,\theta^*}(\mathcal{H}) = \sup_{h \in \mathcal{H}}|\psi_{0,\theta^*}(h)|$. 
Suppose that we have available an estimator $\{\psi_{n,\theta^*}(h):h \in \mathcal{H}\}$ for $\{\psi_{0,\theta^*}(h): h \in \mathcal{H}\}$ and a function class $\mathcal{H}$ that satisfy the conditions of Lemma 2.
Such an estimator could be obtained using either of the strategies presented in Section \ref{methods-estimator}.
Consider the plug-in estimator $\Psi_{n,\theta^*}(\mathcal{H}) := \sup_{h \in \mathcal{H}} |\psi_{n,\theta^*}(h)|$.
The continuous mapping theorem implies that  ${n}^{1/2}\sup_{h \in \mathcal{H}}|\psi_{n,\theta^*}(h) - \psi_{0,\theta^*}(h)|$ converges weakly to $\sup_{h \in \mathcal{H}}|\mathbb{G}(h)|$ where $\mathbb{G}$ is the Gaussian process in Lemma 2.
This, in combination with the reverse triangle inequality together imply that
\begin{align*}
\left|\Psi_{n,\theta^*}(\mathcal{H}) - \Psi_{0,\theta^*}(\mathcal{H})\right| = O_P(n^{-1/2}).
\end{align*}
Furthermore, because when the null hypothesis holds, $\psi_{0,\theta^*}(h) = 0$ for all $h$,  $\Psi_{n, \theta^*}(\mathcal{H})$ converges weakly to $\sup_{h \in \mathcal{H}}|\mathbb{G}(h)|$.
Thus, $\Psi_{n, \theta^*}(\mathcal{H})$ is a consistent estimator for $\Psi_{0,\theta^*}(\mathcal{H})$ that has a fully characterizable null limiting distribution.
This makes it possible to construct an asymptotically valid hypothesis test based on the estimator.

While the null limiting distribution of $\Psi_{n,\theta^*}(\mathcal{H})$ can indeed be characterized, a closed form expression may not be available.
It may therefore be necessary to use an approximation.
We use the multiplier bootstrap method presented in \cite{hudson2021inference}, which makes use of the asymptotic linearity of $\psi_{n,\theta^*}(h)$.
We first note that due to the uniform asymptotic linearity of $\psi_{n,\theta^*}$, under the null hypothesis, $\Psi_{n,\theta^*}(\mathcal{H})$ can be expressed as the sum of the supremum of an empirical process and an asymptotically negligible remainder.
That is,
\begin{align}
\Psi_{n, \theta^*}(\mathcal{H}) = \sup_{h \in \mathcal{H}}\left|n^{-1}\sum_{i=1}^n \phi_{0, \theta^*}(O_i ;h) \right| + o_P(n^{-1/2}).
\label{Psi-n-limit}
\end{align}

We can approximate the null distribution of $\Psi_{n,\theta^*}(\mathcal{H})$ as the supremum of a bootstrapped empirical process that attains the same limiting distribution as the empirical process in \eqref{Psi-n-limit}, conditional on the observed data.
For $m = 1,\ldots, M$ and $M$ large, let $\xi^m_{1}, \ldots, \xi^m_{n}$ be \textit{i.i.d.} random variables, independent of $O_1,\ldots,O_n$, with mean zero, unit variance, and $E_0[|\xi^m_i|^{2 + u}] < \infty$ for some $u > 0$, and let $\phi_{n, \theta^*}$ be an estimator for the efficient influence function.
We define the $m$-th bootstrap sample of $\Psi_{n, \theta^*}(\mathcal{H})$ as
\begin{align}
\Psi^m_{n, \theta^*}(\mathcal{H}) = \sup_{h \in \mathcal{H}}\left|n^{-1}\sum_{i=1}^n \xi^m_i\phi_{n,\theta^*}(O_i; h) \right|.
\label{bootstrap-sample}
\end{align}
It is shown in \cite{hudson2021inference} that if $\phi_{n,\theta^*}$ is a consistent estimator for $\phi_{0,\theta^*}$, and if $\mathcal{H}$ is not overly complex, the multiplier bootstrap statistic converges weakly to $\sup_{h \in \mathcal{H}}|\mathbb{G}(h)|$, conditional on $O_1, \ldots, O_n$.
Thus, the distribution of the multiplier bootstrap samples closely approximates the null limiting distribution of $\Psi_{n,\theta^*}(\mathcal{H})$ in the limit of large $n$.

To estimate the efficient influence function, one approach is to use the plug-in estimator
\begin{align}
\phi_{n, \theta^*}(w,a,y;h) = &\left[ \theta_n(a) - \theta^*(a) - n^{-1}\sum_{i=1}^n\left\{\theta_n(A_i) - \theta^*(A_i)\right\}\right]\left\{h(a) - n^{-1}\sum_{i=1}^nh(A_i)\right\} \nonumber
+
\\
 &\left[\frac{n^{-1}\sum_{i=1}^n g_n(a|W_i)}{g_n(a|w)} \left\{y - Q_n(w,a)   \right\} \right]\left\{h(a) - n^{-1}\sum_{i=1}^nh(A_i)\right\} \nonumber
 +
 \\
&n^{-1}\sum_{i=1}^n Q_n(w,A_i)\left\{h(A_i) - n^{-1}\sum_{j=1}^n h(A_j)\right\} - \nonumber
\\
& n^{-1} \sum_{i=1}^n \left\{Y_i - Q_n(W_i,A_i)\right\}Z_{n}(W_i, A_i;h), \label{eif-estimator}
\end{align}
where we recall that $\theta_n: a \mapsto n^{-1}\sum_{i=1}^n Q_{n}(W_i, a)$ is the plug-in estimator for $\theta_0$, and $Z_n$ is as defined in \eqref{Zn}.
Alternatively, we can observe that because $\bar{\theta}_0 = \bar{\theta}^*$ under the null, in \eqref{eif-estimator}, we can replace $\theta^*$ with an estimator of $\theta_0$, such as the plug-in $\theta_n$.
We note that when we substitute $\theta^*$ by $\theta_n$ in \eqref{eif-estimator}, some cancellation occurs, and the first line in the above expression vanishes.
This strategy of replacing $\theta^*$ with an estimator for $\theta_0$ is appealing because the bootstrap approximation of the limiting distribution no longer depends on $\theta^*$, so one can test any hypothesis of the form \eqref{flatnull} using the same bootstrap sample.
This is particularly useful when we are interested in constructing a confidence set for $\bar{\theta}_0$ by inverting our proposed test, as this requires us to test a large collection of hypotheses.

In our presentation so far, we have assumed that the class $\mathcal{H}$ is fixed.
We acknowledge that in practice, fixing a class a priori may be challenging, so data-adaptive approaches for selecting $\mathcal{H}$ may be preferable.
It has been shown by \cite{hudson2021inference} that data-adaptive selection of $\mathcal{H}$ does not affect the type-1 error rate of our proposed test as long as the data-adaptive choice converges to a fixed class.
In Section 4, we propose an approach for data-adaptive selection of $\mathcal{H}$, and we later show in simulations that our approach is asymptotically valid.

%

%
%
%
%
%

\section{Implementation}

\subsection{Construction of $\mathcal{H}$} \label{implement-H}

Recall that it is our objective to construct $\mathcal{H}$ as a model for $h_0$ in \eqref{steepest-h} so that $\Psi_{0, \theta^*}(\mathcal{H})$ can be interpreted as the standard deviation of the difference between the true dose response curve and the candidate null parameter.
As discussed in Section \ref{methods-test}, we can choose a flexible nonparametric model, so long as the model is not overly complex.
In what follows, we describe a practical approach for selecting such a class.
Our approach is similar to that used by \cite{hudson2021inference} to implement their proposed nonparametric score test.

For a positive semidefinite kernel function $K$ from $\mathcal{A} \times \mathcal{A}$ to $\mathbb{R}^+$, let $\mathcal{S}_K$ denote its unique reproducing kernel Hilbert space (RKHS), endowed with the inner product $\langle \cdot, \cdot \rangle_{\mathcal{S}_K}$.
The kernel function $K$ has the eigen-decomposition
\begin{align*}
(a_1,a_2)\mapsto K(a_1, a_2) = \sum_{d=1}^\infty \gamma_d \eta_d(a_1) \eta_d(a_2),
\end{align*}
where the eigenfunctions $\{\eta_1,\eta_2,\ldots\}$ are orthogonal with respect to the RKHS inner product $\langle \cdot, \cdot \rangle_{\mathcal{S}_K}$, and $0 \leq \gamma_1\leq \gamma_2\leq\ldots$ are the eigenvalues.
Any function $s$ in the RKHS can be expressed as a linear combination of the eigenfunctions.
That is, there exist coefficients $c_1, c_2, \ldots$ such that $s(a) = \sum_{d=1}^\infty c_d \eta_d(a)$ for all $a$.
The roughness of $s$ can be measured by the RKHS norm as
\begin{align*}
J(s) := \langle s, s \rangle_{\mathcal{S}_K} = \sum_{d=1}^\infty\frac{c_d^2}{\gamma_d},
\end{align*}
with higher values of $J(s)$ corresponding to greater roughness. 
We construct $\mathcal{H}$ as a subset of functions in $\mathcal{S}_K$ with bounded roughness and unit variance.
That is
\begin{align*}
\mathcal{H}_\kappa := \left\{h = \textstyle\sum_{d=1}^\infty c_d \eta_d: c_1, c_2, \ldots \in\mathbb{R}, J(h) \leq \kappa, \text{Var}_n(h(A)) = 1\right\},
\end{align*}
where $\text{Var}_n(h(A))$ is the empirical variance of $h(A)$, and $\kappa > 0$ is a tuning parameter.
To facilitate computation, we truncate the eigenbasis at some large level $D$.

In our implementation, we select $\mathcal{S}_K$ as the second-order Sobolev space on $[0,1]$, which can be defined as an RKHS endowed with the inner product $(s_1,s_2)\mapsto\langle s_1, s_2 \rangle_{\mathcal{S}_K} = \int_{0}^1 \ddot{s}_1(u) \ddot{s}_2(u) du$, where $\ddot{s}$ denotes the second derivative of any given function $s$.
In this case, the eigenfunctions and eigenvalues are available in closed form and can be expressed as 
\begin{align*}
\eta_{2d - 1}:a\mapsto\sqrt{2}\cos\left(2 \pi d a \right) ,\quad \eta_{2d}:a\mapsto\sqrt{2} \sin(2 \pi d a), \quad
\gamma_{2d - 1} = \gamma_{2d} = (2\pi d)^{-4},
\end{align*}for $d=1,2,\ldots$ \citep{wahba1990spline}.

We conclude by discussing selection of the tuning parameter $\kappa$.
Our goal is to select $\kappa$ large enough so that $\mathcal{H}_\kappa$ contains a good approximation of a $h_0$.
Suppose $h_0$ belongs in the RKHS $\mathcal{S}_K$.
With prior knowledge on $h_0$ of available, a natural choice would be to set $\kappa = \kappa_0$, where we define
\begin{align*}
\kappa_0 = J(h_0) = \frac{J(\theta_0 - \theta^*)}{\text{Var}_0(\theta_0(A) - \theta^*(A))}.
\end{align*}
Because $\kappa_0$ is typically unknown as it depends $\theta_0 - \theta^*$, we may in practice rely upon an estimate.

We propose to use a simple plug-in estimator for $\kappa_0$.
Consider the following transformation of the observed data:
\begin{align*}
f_n(O_i) = n^{-1} \sum_{j=1}^n Q_n(W_j, A_i) + \frac{n^{-1}\sum_{j=1}^ng_n(A_i|W_j)}{g_n(A_i|W_i)}\{Y_i - Q_n(W_i, A_i)\} - \theta^*(A_i).
\end{align*}
It is shown in \cite{kennedy2017non}  that one can consistently estimate $\theta_0 - \theta^*$ by regressing $f_n(O_i)$ on $A$.
We estimate $\theta_0 - \theta^*$ as $c_{0,n} + \sum_{d=1}^D c_{d,n} \eta_{d}$, where the coefficients are the minimizers of a penalized least squares loss, namely
\begin{align}
c_{0,n}, c_{1,n}, \ldots, c_{d,n} = \underset{c_0,c_1, \ldots, c_d \in \mathbb{R}}{\text{arg min}}\, n^{-1}\sum\left\{f_n(O_i) - c_{0} - \sum_{d=1}^D c_d\eta_d(A_i) \right\}^2 + \lambda \sum_{d=1}^D \frac{c_d^2}{\gamma_d},
\label{pen-least-squares}
\end{align}
where $\lambda > 0$ is a tuning parameter.
The penalty term in \eqref{pen-least-squares} controls the RKHS norm of the resulting estimate, with smaller values of $\lambda$ corresponding to a less smooth estimate.
To select $\lambda$, we perform cross-validation for a large set of candidate values, and we choose the largest candidate for which the cross-validation error is within one standard error of the minimum cross-validation error.
This strategy provides a parsimonious estimate of $\theta_0 - \theta^*$ that fits the observed data well.
In practice, the resulting estimate will often be less rough than $\theta_0 - \theta^*$ \citep{hastie2009elements}.
Finally, we estimate $h_0$ as
\begin{align*}
h_n = \frac{\sum_{d=1}^D c_{d,n} \left\{\eta_j - n^{-1}\sum_{i=1}^n \eta_j(A_i)\right\}}{\text{Var}^{1/2}_n\left(\sum_{d=1}^D c_{d,n} \eta_d(A)\right)},
\end{align*}
and we estimate $\kappa_0$ as
\begin{align*}
\kappa_n = J(h_n) = \frac{\sum_{d=1}^D \frac{c^2_{d,n}}{\gamma_d}}{\text{Var}_n\left(\sum_{d=1}^D c_{d,n} \eta_d(A)\right)}.
\end{align*}

\subsection{Calculation of $\Psi_{n,\theta^*}(\mathcal{H})$} \label{implement-teststat}

We now describe how to calculate $\psi_{n, \theta^*}(h)$ and $\Psi_{n,\theta^*}(\mathcal{H})$.
It is first necessary to estimate the nuisance parameters $Q_0$ and $g_0$.
The conditional mean $Q_0$ can be estimated using any of a wide variety of flexible nonparametric estimators, such as artificial neural networks \citep{barron1989statistical}, the highly adaptive lasso \citep{benkeser2016highly}, or the Super Learner \citep{van2007super}. 
In this work, we use the highly adaptive lasso, which is implemented in the publicly-available R package \texttt{hal9001}.

To construct a nonparametric estimator for the conditional density function $g_0$, we first observe that $g_0$ can be approximated by a conditional mean function.
Let $\pi$ be a non-negative and symmetric function from $\mathbb{R}$ to $\mathbb{R}^+$ for which $\int_{\mathbb{R}} \pi(u) du = 1$.
We define $g_{0,r}$ as
\begin{align*}
g_{0,r}(a|w) = E_0\left[r^{-1}\pi\left(r^{-1}\{A - a\}\right)\bigg|W = w\right],
\end{align*}
where $r > 0$ is a bandwidth.
It can be shown that $g_{0,r}(a|w)$ tends to $g_0(a|w)$ as $r$ tends to zero.
It is therefore sensible to estimate $g_0(a|w)$ using an estimator for the conditional mean $g_{0,r}(a|w)$ for sufficiently small $r$.
We treat the bandwidth $r$ as a tuning parameter that modulates the smoothness of the conditional density estimate in $a$, with smaller $r$ corresponding to lesser smoothness. 
This estimator can be viewed as a generalization of the kernel density estimator for learning a marginal density function.

For a given bandwidth $r$ and a fine grid of fixed points $a_1 < a_2 < \ldots$, one can estimate each $g_{0,r}(a_j| \cdot)$ using a flexible nonparametric estimator for the conditional mean.
One can then obtain estimates at intermediate points via linear interpolation.
To ensure that the conditional density estimate is non-negative, we fit a flexible nonparametric model for $\log g_{0,r}$ using the highly adaptive lasso and transform the resulting model fit, similarly as one would estimate a conditional mean in a generalized linear model with a log link.
The bandwidth $r$ can be selected by performing cross-validation using the log loss function. 
We note that performing cross-validation can be very slow as estimating $g_{0,r}(a_j| \cdot)$ at a large number of grid points for several different choices of bandwidth can be computationally intensive.
An alternative strategy that we suggest is to consider a kernel density estimator for the marginal density of $A$ and to select $r$ as the bandwidth for the kernel density estimator that minimizes the cross-validation error for the marginal density.
We expect this approach to perform reasonably well as long as the conditional density of $A$ at any given $w$ is not much less smooth than the marginal density.

We now discuss calculation of the one-step and TML estimators for $\psi_{0, \theta^*}(h)$, given that estimators $Q_n$ and $g_n$ for $Q_0$ and $g_0$ are available.
One can calculate the one-step estimator at $h = \eta_d$ as
\begin{align*}
\psi^{\mathrm{I}}_{n,\theta^*}(\eta_d) = &n^{-1}\sum_{i=1}^n\left[\theta_n(A_i) - \theta^*(A_i) - n^{-1}\sum_{k=1}^n\left\{\theta_n(A_k) - \theta^*(A_k)\right\}\right]
\left\{\eta_{j}(A_i) - n^{-1}\sum_{k=1}^n \eta_d(A_k)\right\} + 
\\
&n^{-1}\sum_{i=1}^n\left[\left\{\frac{n^{-1}\sum_{k=1}^n g_n(A_i|W_k)}{g_n(A_i|W_i)}\right\}\left\{Y_i - Q_n(W_i, A_i)\right\}  \right]
\left\{\eta_{j}(A_i) - n^{-1}\sum_{k=1}^n \eta_d(A_k)\right\},
\end{align*}
where, as in Section 3.1, $\theta_n: a \mapsto n^{-1}\sum_{i=1}^n Q_n(a, W_i)$ is the plug-in estimator for the dose response function.
Because the one-step estimator is linear in $h$, it easy to see that for any $h = \sum_{j} c_d\eta_d$, $\psi^{\mathrm{I}}_n(h)$ can be expressed as
\begin{align*}
\psi^{\mathrm{I}}_n(h) = \sum_{d=1}^D c_d\psi^{\mathrm{I}}_{n}(\eta_d).
\end{align*}

Computing the TML estimator is more involved than computing the one-step estimator as we need to calculate the TML update $\tilde{Q}_n$ of the initial conditional mean estimator $Q_n$. 
Recall from Section 3.1 that we take $\tilde{Q}_n$ as the minimizer of the squared error loss along the parametric working model $Q_{n,\beta}$ in \eqref{ulfsm}.
For a small $\epsilon > 0$ and a positive integer $B$, we approximate $Q_{n,\beta}$ at $\beta = B\epsilon$ as
\begin{align*}
Q_{n, B\epsilon}(w,a) = Q_{n}(w,a) + \epsilon\sum_{b=1}^B Z_{n}(a, w;h_{n, b\epsilon}),
\end{align*}
where we define $h_{n, b\epsilon}$ as
\begin{align}
h_{n, b\epsilon} := \underset{h \in \mathcal{H}_\kappa}{\text{arg max}}\, n^{-1} \sum_{i=1}^n \left\{Y_i - Q_{n, \epsilon(b-1)}(A_i, W_i)  \right\} Z_{n}(A_i, W_i;h).
\label{hn-b-epsilon}
\end{align}
The optimization problem in \eqref{hn-b-epsilon} is a quadratically constrained quadratic program and can be solved be using publicly available software such as the \texttt{CVXR} package in R \citep{fu2017cvxr}.
One can observe that because $\mathcal{H}_\kappa$ is symmetric in the sense that $h \in \mathcal{H}_\kappa$ implies $-h \in \mathcal{H}_\kappa$, the derivative of the squared error loss, 
\begin{align}
D_n(\beta) := - n^{-1} \sum_{i=1}^n \left\{Y_i - Q_{n, \beta}(A_i, W_i)  \right\} Z_{i,n}(h_{n,\beta}),
\label{hn-b-epsilon}
\end{align}
is necessarily non-positive.
One can therefore find a near-minimizer of the squared error loss by calculating $Q_{n,B\epsilon}$ for incrementally increasing $B$ until $D_n(B\epsilon)$ is sufficiently small.
We take $\tilde{Q}_n = Q_{B_n \epsilon}$, where $B_n$ satisfies 
\begin{align*}
D_n(B_n \epsilon) = \{n \log(n)\}^{-1/2} \text{Var}^{-1/2}_n\left(\left\{Y - Q_n(W,A)\right\}Z_n(W,A;h_n)\right),
\end{align*}
where $h_n$ is the estimator for $h_0$ described in Section \ref{implement-H}.
This choice of $B_n$ ensures that $D_n(B_n \epsilon)$ approaches zero at a rate faster than $n^{-1/2}$, which is a key condition for establishing asymptotic linearity of the TML estimator.
Additionally, choosing $B_n$ so that $D_n(B_n \epsilon)$ is not much smaller than necessary and tends to zero at only a slightly faster rate than $n^{1/2}$ helps to prevent $\tilde{Q}_{n}$ from being an overfitted estimator of $Q_0$.

Now, for any $\eta_d$, the TML estimator of $\psi_{0, \theta^*}(\eta_d)$ can be expressed as
\begin{align*}
\psi^{\mathrm{II}}_{n, \theta^*}(\eta_d) = n^{-1}\sum_{i=1}^n\left[\theta_{\tilde{P}_n}(A_i) - \theta^*(A_i) - n^{-1}\sum_{k=1}^n\left\{\theta_{\tilde{P}_n}(A_k) - \theta^*(A_k)\right\} \right]\eta_d(A_i),
\end{align*}
where we recall that $\theta_{\tilde{P}_n}: a \mapsto n^{-1}\sum_{i=1}^n \tilde{Q}_n(a, W_i)$ is the updated TML estimator for the dose response function.
Because the TML estimator is linear in $h$, for any $h = \sum_{d=1}^D c_d \eta_d$, we have
\begin{align*}
\psi_{n,\theta^*}^{\mathrm{II}}(h) = \sum_{d=1}^D c_d \psi^{\mathrm{II}}_{n, \theta^*}(\eta_d).
\end{align*}

Having described how to compute the one-step and TML estimators for $\psi_{0, \theta^*}(h)$, we now discuss how to calculate $\Psi_{n, \theta^*}(\mathcal{H}_\kappa)$.
Observe that for $h = \sum_{d=1}^D c_d\eta_d$, the one-step and TML estimators are linear in the coefficient vector $\mathbf{c} = (c_1, \ldots, c_d)^\top$ and can be expressed as $U(\theta^*)^\top\mathbf{c}$, where $U(\theta^*)$ is a $D$-dimensional vector for which the $d$-th element contains an estimator for $\psi_{0,\theta^*}(\eta_d)$.
Let $V$ be a $D \times D$ matrix where element $(d_1, d_2)$ is
\begin{align*}
V_{d_1,d_2} = n^{-1}\sum_{i=1}^n \left\{ \eta_{d_1}(A_i) - n^{-1} \sum_{j=1}^n \eta_{d_1}(A_j) \right\}\left\{ \eta_{d_2}(A_i) - n^{-1} \sum_{j=1}^n \eta_{d_2}(A_j) \right\}
\end{align*}
so that  the empirical variance of $h(A)$  is $\text{Var}_n(h(A)) = \mathbf{c}^\top V \mathbf{c}$, and let $\Gamma = \text{diag}(\gamma_1^{-1}, \ldots, \gamma_d^{-1})$, where $\gamma_1,\ldots,\gamma_d$ are the eigenvalues for the kernel $K$.
We can express $\Psi_{n,\theta^*}(\mathcal{H}_\kappa)$ as $U(\theta^*)^\top \mathbf{c}_n$, where $\mathbf{c}_n$ is defined as
\begin{align}
\mathbf{c}_n := \text{arg max}\left\{ U(\theta^*)^\top \mathbf{c}: \mathbf{c}^\top V \mathbf{c} = 1, \mathbf{c}^\top \Gamma \mathbf{c} \leq \gamma \right\}.
\label{constrained-opt}
\end{align}
The Karush-Kuhn-Tucker conditions for the optimization problem in \eqref{constrained-opt}  imply that $\mathbf{c}_n$ is the solution to
\begin{align*}
U(\theta^*) - \lambda_1  \left(V + \lambda_2\Gamma\right) \mathbf{c} = 0,
\end{align*}
where $\lambda_1 > 0$ and $\lambda_2 > 0$ are chosen so that the constraints are satisfied.
With some algebra, one can show that
\begin{align*}
\mathbf{c}_n  = \lambda_{1,n}^{-1}(V + \lambda_{2,n}\Gamma)^{-1}U(\theta^*),
\end{align*}
where $\lambda_{2,n}$ satisfies
\begin{align}
\frac{U(\theta^*)^\top(V + \lambda_{2,n} \Gamma)^{-1} \Gamma (V + \lambda_{2,n}\Gamma)^{-1}U(\theta^*)}{U(\theta^*)^\top(V + \lambda_{2,n} \Gamma)^{-1} V (V + \lambda_{2,n}\Gamma)^{-1}U(\theta^*)} = \gamma,
\label{lam1}
\end{align}
and
\begin{align}
\lambda_{1,n} = \left\{ U(\theta^*)^\top(V + \lambda_{2,n} \Gamma)^{-1}V(V + \lambda_{2,n} \Gamma)^{-1} U(\theta^*) \right\}^{1/2}.
\label{lam2}
\end{align}
Finally, we can express $\Psi_n(\mathcal{H}_\kappa)$ as
\begin{align*}
\Psi_{n,\theta^*}(\mathcal{H}_\kappa) = \lambda_{1,n}^{-1}U(\theta^*)^\top(V + \lambda_{2,n}\Gamma)^{-1}U(\theta^*).
\end{align*}

\subsection{Bootstrap Approximation of the Null Limiting Distribution}

We now discuss how to draw multiplier bootstrap samples to estimate the null limiting distribution of $\Psi_{n, \theta^*}(\mathcal{H}_\kappa)$.
For $m = 1, \ldots, M$ and $M$ large, we draw $\xi^{(m)}_1, \dots, \xi^{(m)}_n$ as independent standard normal random variables.
Let $\bar{\xi}^{(m)} = n^{-1}\sum_{i=1}^n \xi_i^{(m)}$, and let $U^{(m)}$ be a $D-$dimensional vector with $d$-th element
\begin{align*}
U_d^{(m)} = n^{-1}\sum_{i=1}^n\left(\xi_i^{(m)} - \bar{\xi}^{(m)} \right) \phi_{n, \theta^*}(O_i;\eta_d),
\end{align*}
where we recall that $\phi_{n, \theta^*}$ is the plug-in estimator for the efficient influence function in \eqref{eif-estimator} (and as noted in Section 3.2, we may wish to replace $\theta^*$ with an estimator for $\theta_0$).
The $m$-th sample from the multiplier bootstrap estimate of the null distribution (see \eqref{bootstrap-sample}) can be calculated as 
\begin{align*}
\Psi^{(m)}_{n, \theta^*}(\mathcal{H}_\kappa) = \sup_{\mathbf{c} \in \mathbf{R}^d}\left\{\left(U^{(m)}\right)^\top \mathbf{c}: \mathbf{c}^\top \Gamma \mathbf{c} \leq \gamma, \mathbf{c}^\top V \mathbf{c} = 1\right\}. 
\end{align*}
The above optimization problem can be solved using the same routine described in Section \ref{implement-teststat}, simply replacing $U$ with $U^{(m)}$.
Finally, for a realization $t$ of $\Psi_{n, \theta^*}(\mathcal{H}_\kappa)$, a bootstrap p-value can be calculated as
\begin{align*}
\rho_M(t) = M^{-1}\sum_{m=1}^M \mathds{1}\left(\Psi^{(m)}_{n, \theta^*}(\mathcal{H}_\kappa) > t \right).
\end{align*}

\subsection{Confidence Band Construction}

In this section, we discuss how to visualize the confidence set for $\bar{\theta}_0$ obtained by inverting our proposed hypothesis test.
Recall from Section \ref{overview-summary} that we propose to report the smallest band that contains all functions belonging to the confidence set.
This confidence band is defined as $C_n^\alpha$, and its form is provided in \eqref{conf-band}.

We first need to construct a function class $\Theta$ that contains a collection of candidate values for $\theta_0$.
While $\Theta$ can be a rich class, it cannot be entirely unrestricted.
In fact, if $\Theta$ is too large, the confidence band $C_n^\alpha$ can possibly have infinite width.
To see this, note that the confidence set $\mathcal{C}_n^\alpha$ in \eqref{conf-set} contains a set of functions $\theta$ for which $\Psi_{n,\theta}(\mathcal{H})$ is close to zero.
It is possible to construct $\theta$ so that at any given $a$, $\theta(a)$ takes an arbitrarily large positive or negative value, but $\Psi_{n,\theta}(\mathcal{H}) = 0$.
For instance, if $\Psi_{n,\theta}(\mathcal{H})$ is constructed using targeted minimum loss-based estimation, this could be achieved by setting $\theta(A_i) = \theta_{\tilde{P}_n}(A_i)$ for $i = 1,\ldots, n$ and allowing $\theta$ to take \textit{any value} at points where no data are observed.
We would encounter the same issue if we instead used the one-step estimator.
By selecting $\Theta$ as, e.g., a class of smooth functions, we are able to avoid this problem.
On the other hand, we note that if $\Theta$ is not large enough to contain $\theta_0$, $\mathcal{C}_n^\alpha$ is not guaranteed to achieve the nominal coverage rate.
Given these considerations, we suggest selecting $\Theta$ as a class of functions that is no less smooth than a reasonable approximation of $\theta_0$.
In our implementation, we construct $\Theta$ as a subset of functions belonging to an RKHS for which the RKHS norm bounded above by  a constant. 
That is, we take $\Theta = \Theta_\nu$, where
\begin{align*}
\Theta_{\nu} := \left\{\theta = \sum_{d=1}^D c_d \eta_d : \sum_{d=1}^D \frac{c_d^2}{\gamma_d} <  \nu\right\}
\end{align*}
for $\nu > 0$.
We propose to set $\nu$ as the RKHS norm of a consistent estimate of $\theta_0$, which could be obtained using the method described in Section \ref{implement-H}.

Let $t^*_{1-\alpha}$ be the $1-\alpha$ quantile of the null limiting distribution of $n^{1/2}\Psi_{n,\theta^*}(\mathcal{H}_\kappa)$ for a fixed $\kappa$.
Given the above construction of $\Theta$, the confidence band $C_n^\alpha$ takes the following form at any given point $a_0$:
\begin{align}
C^\alpha_n(a_0) = \Bigg[&\inf \left\{\sum_d c_d\left\{\eta_d(a_0) - n^{-1}\sum_{i=1}^n \eta_d(A_i)  \right\} : \sum_{d=1}^D \frac{c_d^2}{\gamma_d} \leq \nu, \Psi_{n, \sum_d c_d \eta_d}(\mathcal{H}_\kappa) \leq n^{-1/2}t^*_{1-\alpha}  \right\}, \nonumber
\\
&\sup \left\{\sum_d c_d\left\{\eta_d(a_0) - n^{-1}\sum_{i=1}^n \eta_d(A_i)  \right\} : \sum_{d=1}^D \frac{c_d^2}{\gamma_d} \leq \nu, \Psi_{n, \sum_d c_d \eta_d}(\mathcal{H}_\kappa) \leq n^{-1/2}t^*_{1-\alpha}  \right\}\Bigg].
\label{conf-band-numerical}
\end{align}

The optimization problems in \eqref{conf-band-numerical} are challenging to solve because $\Psi_{n, \sum_{d} c_d \eta_d}(\mathcal{H})$ does not have a closed form expression in the coefficients $c_1, \ldots, c_D$.
Recall from Section \ref{implement-teststat} that we can write
\begin{align*}
\Psi_{n,\sum_d c_d \eta_d}(\mathcal{H}_\kappa) = \lambda_{1,n}^{-1}U\left(\sum_{d=1}^D c_d\eta_d \right)^\top\left(V + \lambda_{2,n}\Gamma\right)^{-1}U\left(\sum_{d = 1}^D c_d \eta_d\right),
\end{align*}
where $\lambda_{1,n}$ and $\lambda_{2,n}$ are constants that depend on $U\left(\sum_{d=1}^D c_d\eta_d \right)$.
By instead treating $\lambda_{1,n}$ and $\lambda_{2,n}$ as fixed, we are able to obtain a closed form approximation of $\Psi_{n,\theta^*}(\mathcal{H}_\kappa)$.
When $\Psi_{n,\theta^*}(\mathcal{H}_\kappa)$ is constructed using either TML or one-step estimation, $U\left(\sum_{d=1}^D c_d\eta_d \right)$ is linear in the coefficients.
As a result, when the closed form approximation of the test statistic is used, the optimzation problem in \eqref{conf-band-numerical} becomes a quadrtically constrained quadratic program.
As noted previously, this type of problem can be solved using publicly available software such as the \texttt{CVXR} package in R.

We conclude by discussing how to select the tuning parameters $\kappa$, $\lambda_{1}$, and $\lambda_2$.
The choice of $\kappa$ should have no bearing on the asymptotic coverage of the confidence set, though it may affect the confidence band's width.
In order for the confidence band to have optimal width, we need to select $\kappa$ as to maximize the power to reject any null hypothesis $H_0: \bar{\theta}_0 = \bar{\theta}^*_0$.
Though the optimal choice of $\kappa$ generally depends on the specific null hypothesis being tested, the optimization problem in \eqref{conf-band-numerical} would become complicated if $\mathcal{H}_\kappa$ was not fixed.
For computational ease, we fix $\kappa$ as a single value that is large enough so that we have reasonable power to reject a large set of null hypotheses.
We set $\kappa$ as an estimate for $J(\theta_0)/\text{Var}_0(\theta_0(A))$, which can be obtained using the approach described in Section \ref{implement-H}, so that we are well powered against nearly flat nulls when $\theta_0$ is not very flat.
Finally, we pick $\lambda_{1,n}$ and $\lambda_{2,n}$ to satisfy \eqref{lam1} and \eqref{lam2} with $\theta^* = 0$.
For this choice of $\lambda_{1,n}$ and $\lambda_{2,n}$, the closed form approximation of $\Psi_{n, \theta^*}(\mathcal{H})$ will be fairly accurate when $\theta^*$ is nearly flat.
Though $\lambda_{1,n}$ and $\lambda_{2,n}$ are data-dependent, asymptotic coverage will be unaffected if $\lambda_{1,n}$ and $\lambda_{2,n}$ converge to fixed constants.

\section{Simulations Study}


\subsection{Simulation Setting}

We begin by describing our approach for generating synthetic data sets.
We first generate $W_1, \ldots, W_n$ as independent bivariate normal random vectors with mean zero, unit variance, and correlation $\frac{1}{2}$.
Given $W$, we then draw $A_1, ..., A_n$ from a conditional distribution with density function
\begin{align*}
g_0(a|w) = \frac{\expit(\zeta(w)a)}{\int_{-1}^1 \expit(\zeta(w)a)da}\mathds{1}(-1 \leq a \leq 1),
\end{align*}
where we define $\zeta$ as
\begin{align*}
\zeta(w) =  3\left\{\text{expit}(w_1 + w_2) - \frac{1}{2}\right\}.
\end{align*}
Random variables with the above conditional density can be generated via the inverse cumulative distribution function method.

\begin{figure}[!h]
\center
\includegraphics[scale=.65]{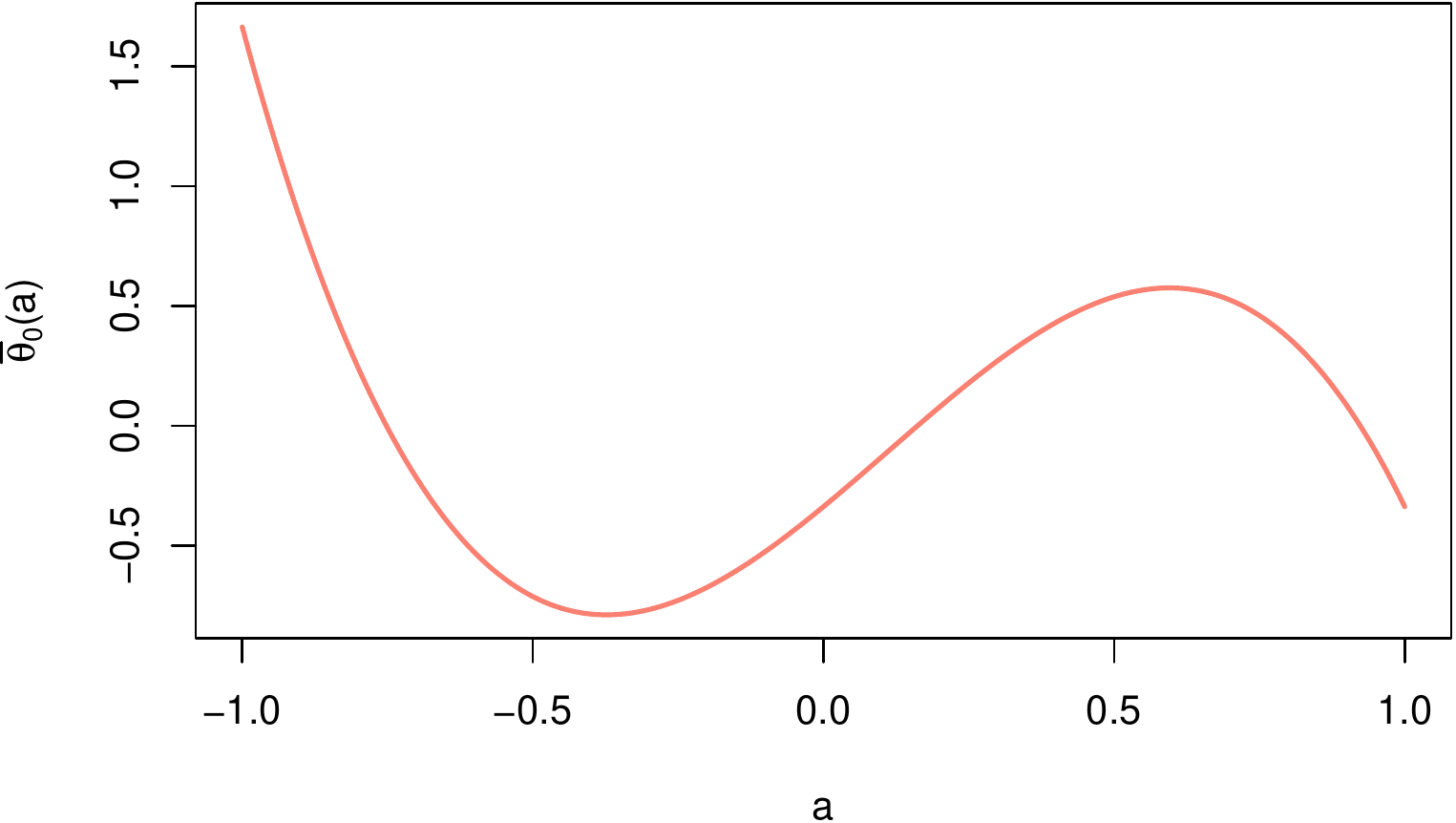}
\caption{Plot of the dose response function under the data-generating mechanism used in the simulation study.}
\label{fig:drf-graph}
\end{figure}

We generate the outcome $Y$ under the following settings.
\\
\textit{Setting 1:}
\\
In the first setting, we construct the conditional distribution of the outcome given the exposure and covariates so that the centered dose response curve is zero.
We draw $Y$ from the model 
\begin{align*}
Y = -\zeta(W)\left\{1 - \frac{A}{2}\right\} + \epsilon,
\end{align*}
where $\epsilon$ is a uniform random variable on $[-2, 2]$.
Because $\zeta(W)$ has mean zero, it can be seen that $E_0[Q_0(W,a)] = 0$ as desired.
\\
\textit{Setting 2:}
\\
In the second setting, we consider the case where $\bar{\theta}_0$ is non-zero.
We construct a model for $Y$ so that the dose response function is
\begin{align*}
\theta_0: a \mapsto \frac{2a + a^2 - a^3}{2}.
\end{align*}
A plot of the dose response function is provided in Figure \ref{fig:drf-graph}.
We generate $Y$ as
\begin{align*}
Y = \theta_0(A) - \zeta(W)\left\{1 - \frac{A}{2}\right\} + \epsilon,
\end{align*}
where $\epsilon$ is again a uniform random variable on $[-2, 2]$.

Under each of the above settings, we use our proposed methodology to perform a test of the null hypothesis that the dose function is flat, i.e., $H_0: \bar{\theta}_0 = 0$.
In the first setting, the null holds, and we would expect our approach to achieve nominal type-1 error control in the limit of large $n$.
Under the second setting, the alternative holds, allowing us to assess the power of our proposed test.
We also examine the behavior of the proposed confidence bands in this setting.
We assess whether the bands are appropriate in the sense that they roughly capture the shape of the unknown dose response function.

We study the behavior of the following four variations of our proposal:
\begin{enumerate}
\item A one-step estimator for $\Psi_{0,\theta^*}$ is used, and we set $\kappa$ as the oracle $\kappa_0$.
\item A TML estimator for $\Psi_{0,\theta^*}$ is used, and we set $\kappa$ as the oracle $\kappa_0$.
\item A one-step estimator for $\Psi_{0,\theta^*}$ is used, and we set $\kappa$ as the data-adaptive choice $\kappa_n$.
\item A TML estimator for $\Psi_{0,\theta^*}$ is used, and we set $\kappa$ as the data-adaptive choice $\kappa_n$.
\end{enumerate}
In each case, we use $D = 20$ basis functions.

We compare our proposed hypothesis test with an approach similar to that described in \cite{westling2021nonparametric}, which is based on estimating primitive functions of the dose response function. 
As noted above, their approach can be viewed as a variation of our proposal where we set $\mathcal{H} = \{h(a) = \mathds{1}(a \leq a_0): a_0 \in \mathcal{A}\}$.  
We use our own implementation of this procedure, which differs slightly in that we estimate $\psi_{0,\theta^*}(h)$ using a one-step estimatior, whereas \cite{westling2021nonparametric} uses a cross-fitted estimator.
We apply each of the above methods to 500 synthetic data sets for $n \in \{100, 200, \ldots, 500\}$.

\subsection{Simulation Results}

\begin{figure}[!h]
\center
\includegraphics[scale=.875]{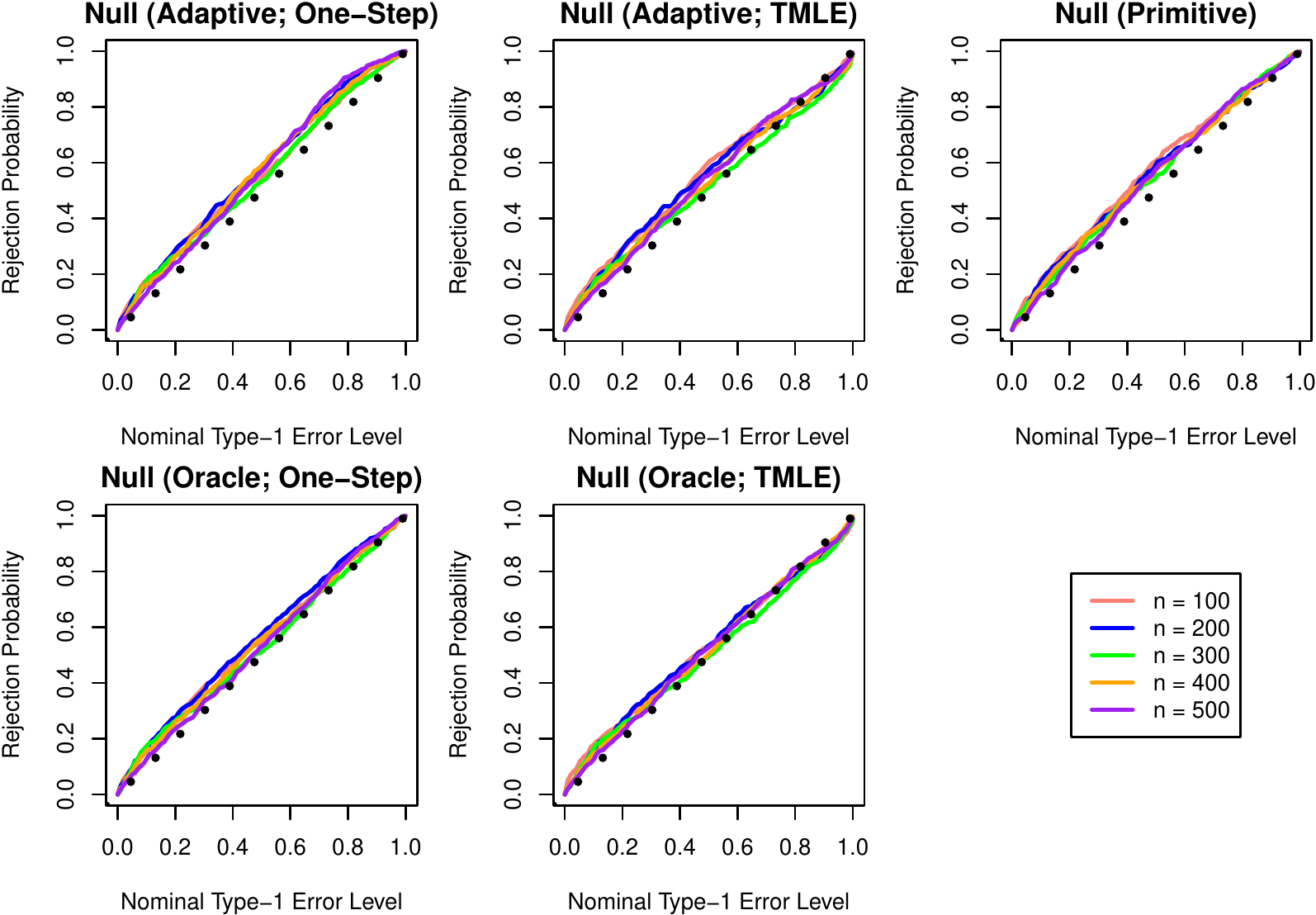}
\caption{Monte Carlo estimates of the the empirical distribution of the p-values under the flat null.}
\label{fig:sims-null}
\end{figure}

Figure \ref{fig:sims-null} shows the Monte Carlo estimate of the distribution function for the p-values produced from each method under Setting 1, where the flat null holds.
When the type-1 error rate is well-controlled for any significance level $\alpha$, the distribution function should be linear.
We find that our approach achieves type-1 error control near the nominal error level when the oracle choice of $\kappa$ is provided, and the data-adaptive choice results in some modest anti-conservatism.
The approach based on estimation of primitive functions also achieves nearly nominal type-1 error control.

\begin{figure}[!h]
\center
\includegraphics[scale=.875]{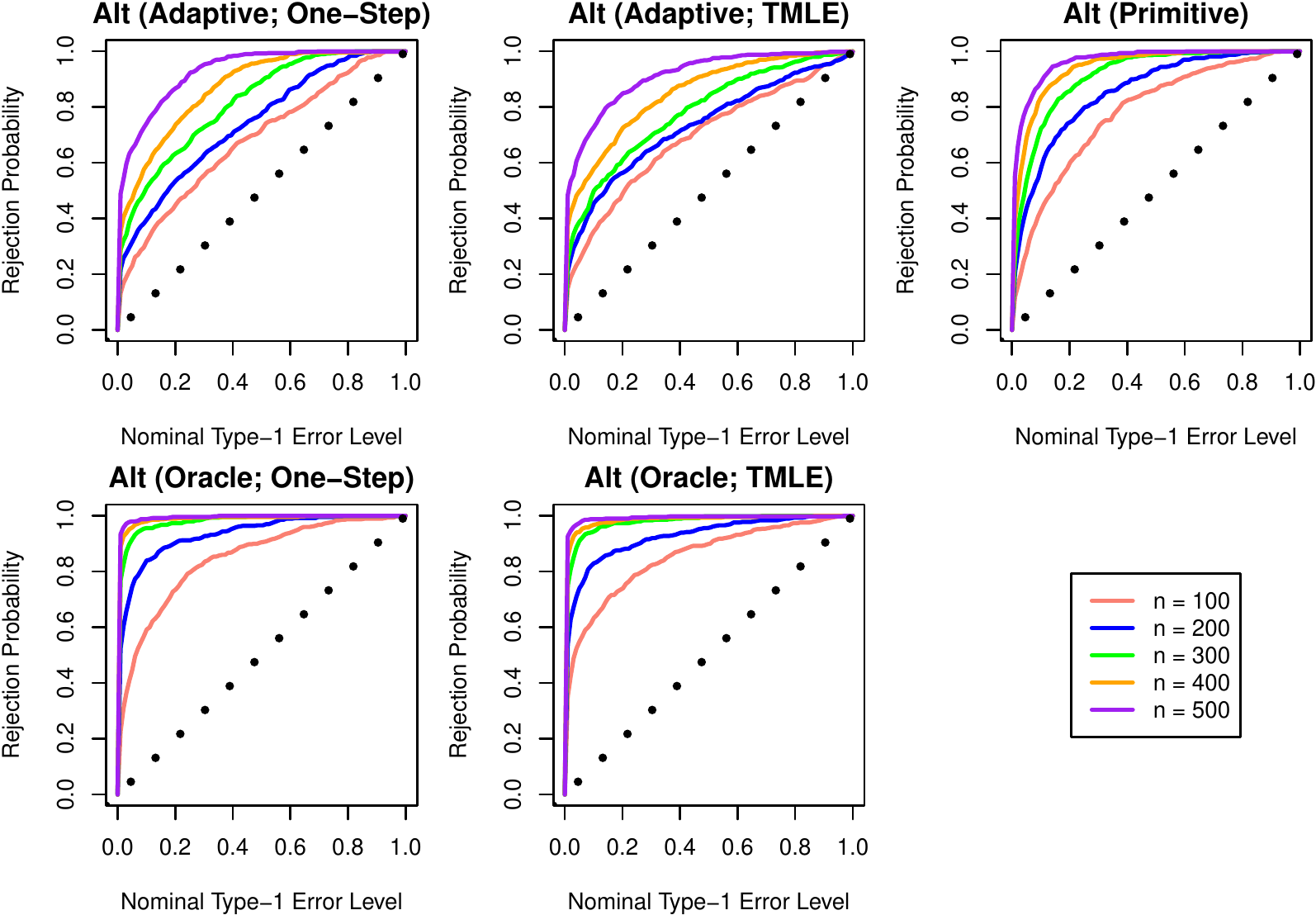}
\caption{Monte Carlo estimates of the empirical distribution of the p-values under for a test against the flat null, under the alternative hypothesis.}
\label{fig:sims-power}
\end{figure}

Figure \ref{fig:sims-power} shows the Monte Carlo estimate of the distribution functions for the p-values under Setting 2, where the alternative holds.
We find that our proposal has high power when the oracle choice of $\kappa$ is supplied, and power declines when a data-adaptive choice is used.
The approach proposed in \cite{westling2021nonparametric} outperforms our approach when $\kappa$ is chosen data-adaptively but performs worse than our approach when the oracle choice is used.
This suggests that making use of known structure on $\theta_0$ can help us improve power to reject some alternative hypotheses, though there is a notable decline in performance when we attempt to learn the structure from the data.

\begin{figure}[!h]
\center
\includegraphics[scale=.875]{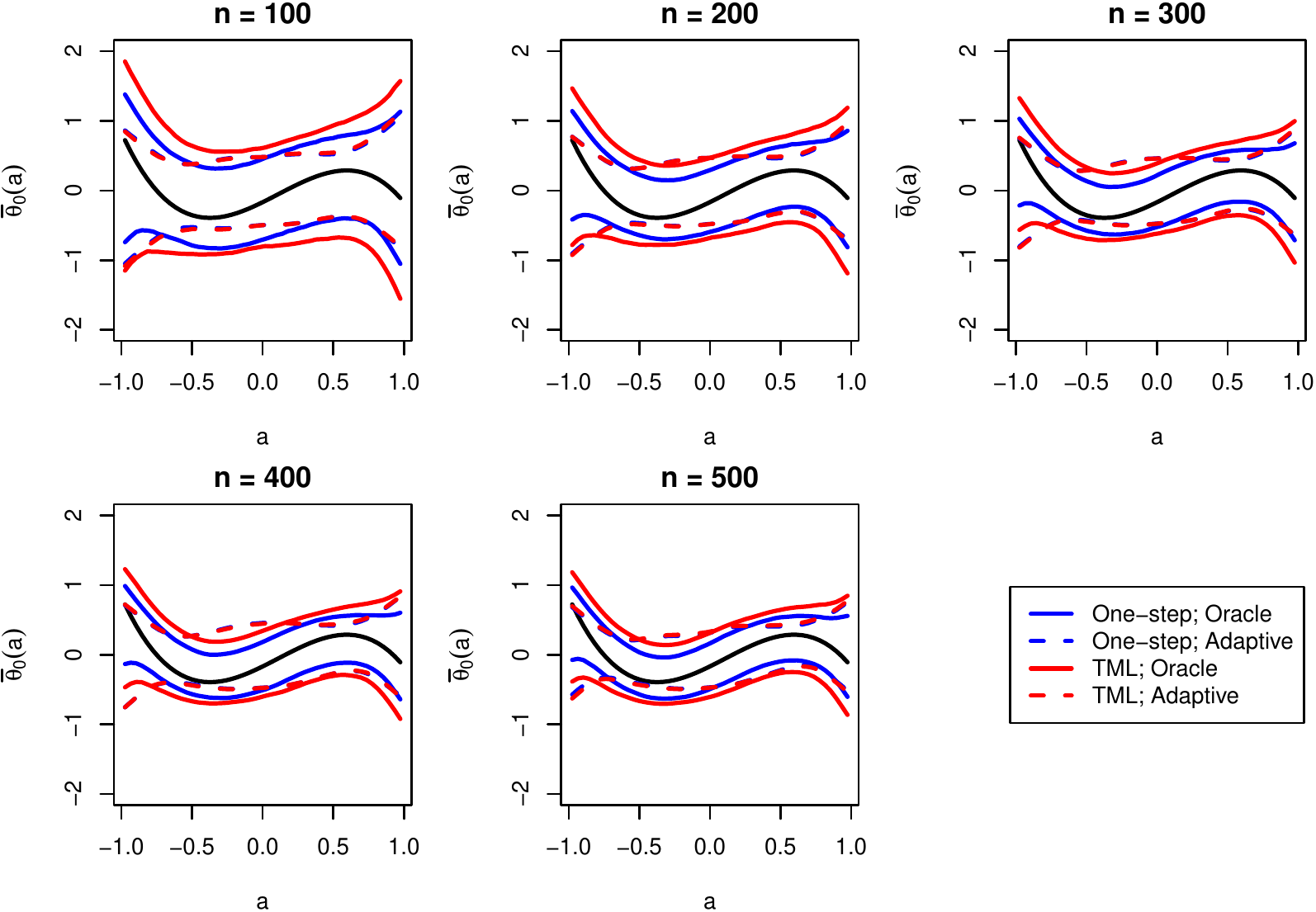}
\caption{Monte Carlo estimates of the median upper and lower limits of the $95\%$ bands obtained using our proposed methodology.}
\label{fig:sims-bands}
\end{figure}

Figure \ref{fig:sims-bands} shows the median upper and lower limits of the confidence bands that were constructed using our proposal.
We find that the confidence bands are able to capture the shape of the dose response curve, and the width of the bands decreases as the sample size grows, as expected.

\section{Data Example}

As an example, we use our method to analyze data from the Adaptive Strategies for Preventing and Treating Lapses of Retention in HIV Care (ADAPT-R) trial \citep{geng2023adaptive}.
ADAPT-R was a sequential multiple assignment randomized trial run in Kenya that studied 
the effectiveness of interventions for optimizing retention in HIV treatment in a population of people living with HIV who initiated care.
In this study, a question of secondary interest is whether the distance a participant must travel to reach the nearest HIV clinic affects their retention in care.
We perform an analysis to address this secondary aim, pooling across the trial's randomized arms.

We conduct our analysis using a sample of 1815 participants from the ADAPT-R trial.
We treat as the exposure of interest the distance from the nearest clinic.
The distribution of the exposure variable is highly skewed.
Approximately 95\% of study participants lived within 20 kilometers of the nearest clinic, and among the remaining participants, distance ranges between 20 and 500 kilometers.
Because these extreme values are fairly rare, there is concern about potential violation of the positivity assumption.
To avoid this issue, we exclude from this analysis participants who lived more than 20 kilometers from the nearest clinic, obtaining a final sample size of 1600.
Our outcome is a binary variable that is equal to one if a patient had neither experienced a lapse in care (defined as missing a scheduled clinic visit by at least 14 days)  nor had unsuppresed HIV viral load one year after initiating care, and is zero otherwise.
A total of 446 study participants experienced a lapse in care or had unsuppresed HIV viral load within one year.
 As our exposure of interest is not randomized, we adjust for the following set of measured baseline variables that may either confound the exposure outcome relationship, or predict the outcome and thus improve efficiency: age, sex assigned at birth, and a wealth index.

\begin{figure}[!h]
\center
\includegraphics[scale=.875]{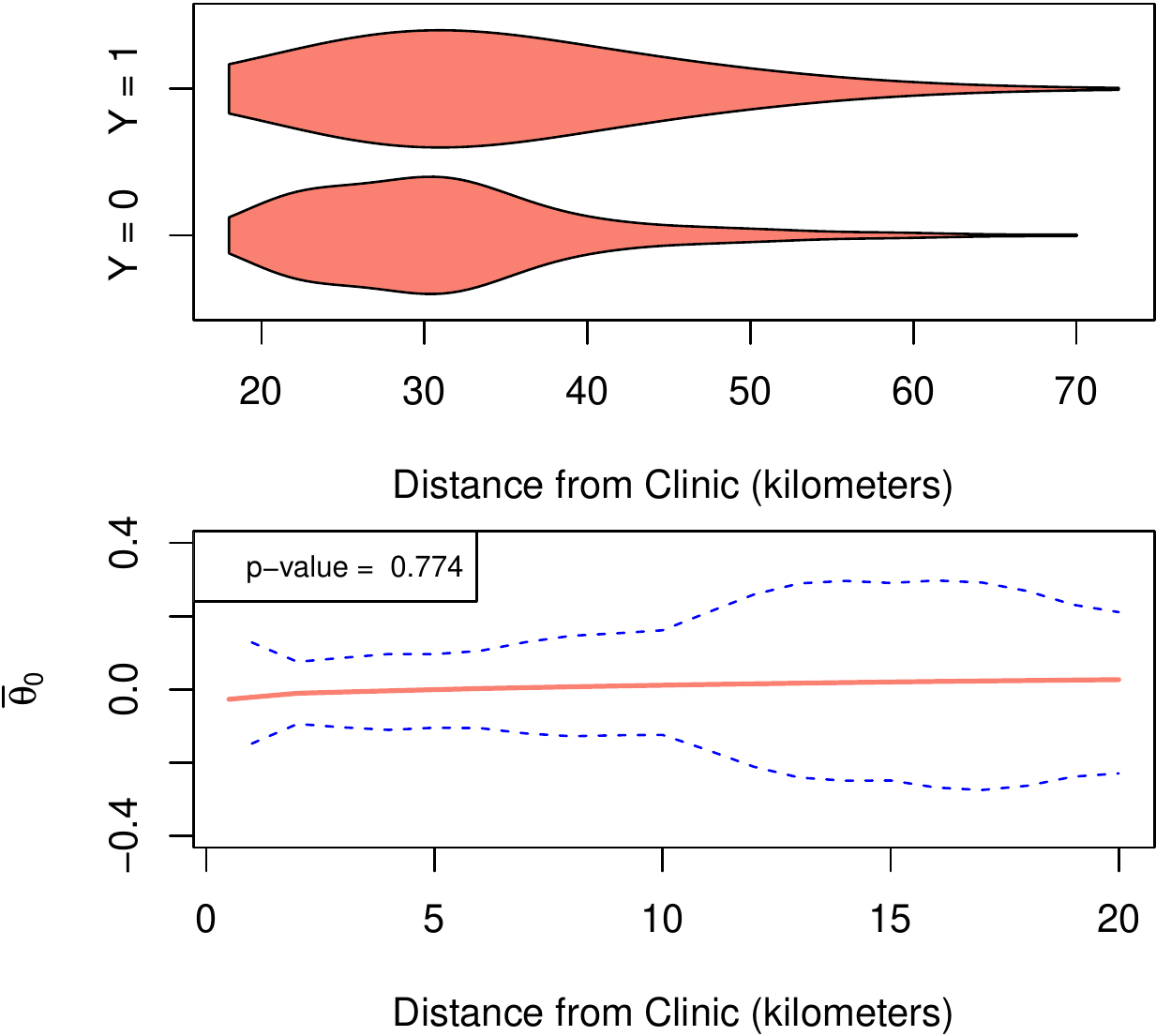}
\caption{ (Top) Violin plot of distribution of the distance to furthest clinic, by status of retention in care. (Bottom) Plug-in estimate of centered dose-response function and 95\% confidence bands.}
\label{fig:data}
\end{figure}

In Figure \ref{fig:data}, we display the marginal distribution of distance to clinic by retention status.
There does not appear to be a strong association between distance and retention, as the marginal distribution is nearly the same in both groups.
To more formally assess the presence of an effect, we apply our method to perform a test of the flat null.
We use the data-adaptive choice of $\kappa$ described in Section \ref{implement-H}, and we use a one-step estimator for $\Psi_{0,\theta^*}(\mathcal{H})$.
Figure \ref{fig:data} shows a plug-in estimate of the centered dose response function, in addition to 95\% confidence bands and a p-value for a test of the flat null.
The dose-response function appears to be nearly flat, and we are unable to reject the flat nulll hypothesis based as our p-value is quite large.
These results suggest that there is not strong evidence to support that distance from clinic has a strong effect on retention in care in people living with HIV in this setting.

\section{Conclusion}

This work provides a novel approach to inference on the causal dose-response function.
We show that, under mild regularity conditions, our nonparametric test achieves type-1 error control near the nominal level and is well-powered against the null.
We also present a computationally tractable method for visualizing confidence sets constructed by inverting our proposed test.
The recent proposal by \cite{westling2021nonparametric} also performs well under weak assumptions, though their work does not present a method for constructing or visualizing confidence sets.
That we introduce a novel approach for constructing confidence sets is a key strength of our work.

The strategy for inference on the dose-response function we describe in this paper can be adapted to address other problems of interest in the causal inference literature.
For instance, one could use our approach to assess for treatment effect heterogeneity by testing the null hypothesis that a conditional average treatment effect curve is flat.

One of the main limitations of this work is that we require a specification for the function class $\mathcal{H}$ for our test to be operational, though this class can be challenging to select in practice.
When $\mathcal{H}$ is specified a-priori and contains $h_0$, our proposal performs very well, but when we attempt to select $\mathcal{H}$ data-adaptively, we suffer a loss in performance.
We note that in some settings selecting $\mathcal{H}$ a priori may be possible.
For instance, if an independent data set is available (e.g., from a closely-related study), one could use this data set to construct $\mathcal{H}$ without looking at the data set they are primarily interested in analyzing.
Alternatively, it is sensible in many settings to assume, without looking at the data, that the dose-response function is monotone.
Therefore, one could consider implementing a version of our procedure where $\mathcal{H}$ is a class of bounded monotone functions.
In future work, we plan to develop improved strategies for tuning parameter selection.

Our proposal also requires that the nuisance parameter estimators are not overly complex.
This condition is somewhat prohibitive and disallows us from using more flexible estimators, such as gradient-boosted trees \citep{friedman2002stochastic}.
To avoid this assumption, one could develop a slightly modified version of our procedure where $\psi_{0, \theta^*}$ is estimated using cross-fitting  \citep{zheng2011cross, chernozhukov2018double}.

\section*{Acknowledgements}

We thank the Family AIDS Care and Education Services program operating HIV services in western Kenya, as well as the patients in these communities and the front-line health care workers in the region.
This research was supported by grants (R01 MH104123, K24 AI134413, and R01 AI074345) from the National Institutes of Health.

\newpage

\bibliography{npscore-drf.bib}

\begin{thebibliography}{}

\bibitem[\protect\citeauthoryear{Barron}{Barron}{1989}]{barron1989statistical}
Barron, A.~R. (1989).
\newblock Statistical properties of artificial neural networks.
\newblock In {\em Proceedings of the 28th IEEE Conference on Decision and
  Control,}, pages 280--285. IEEE.

\bibitem[\protect\citeauthoryear{Benkeser and van~der Laan}{Benkeser and
  van~der Laan}{2016}]{benkeser2016highly}
Benkeser, D. and van~der Laan, M. (2016).
\newblock The highly adaptive lasso estimator.
\newblock In {\em 2016 IEEE international conference on data science and
  advanced analytics (DSAA)}, pages 689--696. IEEE.

\bibitem[\protect\citeauthoryear{Bickel, Klaassen, Ritov, and Wellner}{Bickel
  et~al.}{1998}]{bickel1998efficient}
Bickel, P.~J., Klaassen, C.~A., Ritov, Y., and Wellner, J.~A. (1998).
\newblock {\em Efficient and adaptive estimation for semiparametric models}.
\newblock Springer.

\bibitem[\protect\citeauthoryear{Chernozhukov, Chetverikov, Demirer, Duflo,
  Hansen, Newey, and Robins}{Chernozhukov
  et~al.}{2018}]{chernozhukov2018double}
Chernozhukov, V., Chetverikov, D., Demirer, M., Duflo, E., Hansen, C., Newey,
  W., and Robins, J. (2018).
\newblock {Double/debiased machine learning for treatment and structural
  parameters}.
\newblock {\em The Econometrics Journal} {\bf 21,} C1--C68.

\bibitem[\protect\citeauthoryear{Colangelo and Lee}{Colangelo and
  Lee}{2020}]{colangelo2020double}
Colangelo, K. and Lee, Y.-Y. (2020).
\newblock Double debiased machine learning nonparametric inference with
  continuous treatments.
\newblock {\em arXiv preprint arXiv:2004.03036} .

\bibitem[\protect\citeauthoryear{D{\'\i}az and van~der Laan}{D{\'\i}az and
  van~der Laan}{2013}]{diaz2013targeted}
D{\'\i}az, I. and van~der Laan, M.~J. (2013).
\newblock Targeted data adaptive estimation of the causal dose--response curve.
\newblock {\em Journal of Causal Inference} {\bf 1,} 171--192.

\bibitem[\protect\citeauthoryear{Friedman}{Friedman}{2002}]{friedman2002stochastic}
Friedman, J.~H. (2002).
\newblock Stochastic gradient boosting.
\newblock {\em Computational statistics \& data analysis} {\bf 38,} 367--378.

\bibitem[\protect\citeauthoryear{Fu, Narasimhan, and Boyd}{Fu
  et~al.}{2017}]{fu2017cvxr}
Fu, A., Narasimhan, B., and Boyd, S. (2017).
\newblock {CVXR}: An r package for disciplined convex optimization.
\newblock {\em arXiv preprint arXiv:1711.07582} .

\bibitem[\protect\citeauthoryear{Galvao and Wang}{Galvao and
  Wang}{2015}]{galvao2015uniformly}
Galvao, A.~F. and Wang, L. (2015).
\newblock Uniformly semiparametric efficient estimation of treatment effects
  with a continuous treatment.
\newblock {\em Journal of the American Statistical Association} {\bf 110,}
  1528--1542.

\bibitem[\protect\citeauthoryear{Geng, Odeny, Montoya, Iguna, Kulzer, Adhiambo,
  Eshun-Wilson, Akama, Nyandieka, Guz{\'e}, et~al\mbox{.}}{Geng
  et~al.}{2023}]{geng2023adaptive}
Geng, E.~H., Odeny, T.~A., Montoya, L.~M., Iguna, S., Kulzer, J.~L., Adhiambo,
  H.~F., Eshun-Wilson, I., Akama, E., Nyandieka, E., Guz{\'e}, M.~A., et~al.
  (2023).
\newblock Adaptive strategies for retention in care among persons living with
  hiv.
\newblock {\em NEJM Evidence} {\bf 2,} EVIDoa2200076.

\bibitem[\protect\citeauthoryear{Hastie, Tibshirani, Friedman, and
  Friedman}{Hastie et~al.}{2009}]{hastie2009elements}
Hastie, T., Tibshirani, R., Friedman, J.~H., and Friedman, J.~H. (2009).
\newblock {\em The elements of statistical learning: data mining, inference,
  and prediction}, volume~2.
\newblock Springer.

\bibitem[\protect\citeauthoryear{Hines, Diaz-Ordaz, and Vansteelandt}{Hines
  et~al.}{2021}]{hines2021parameterising}
Hines, O., Diaz-Ordaz, K., and Vansteelandt, S. (2021).
\newblock Parameterising the effect of a continuous exposure using average
  derivative effects.
\newblock {\em arXiv preprint arXiv:2109.13124} .

\bibitem[\protect\citeauthoryear{Hirano and Imbens}{Hirano and
  Imbens}{2004}]{hirano2004propensity}
Hirano, K. and Imbens, G.~W. (2004).
\newblock The propensity score with continuous treatments.
\newblock {\em Applied Bayesian modeling and causal inference from
  incomplete-data perspectives} {\bf 226164,} 73--84.

\bibitem[\protect\citeauthoryear{Hudson, Carone, and Shojaie}{Hudson
  et~al.}{2021}]{hudson2021inference}
Hudson, A., Carone, M., and Shojaie, A. (2021).
\newblock Inference on function-valued parameters using a restricted score
  test.
\newblock {\em arXiv preprint arXiv:2105.06646} .

\bibitem[\protect\citeauthoryear{Imai and Van~Dyk}{Imai and
  Van~Dyk}{2004}]{imai2004causal}
Imai, K. and Van~Dyk, D.~A. (2004).
\newblock Causal inference with general treatment regimes: Generalizing the
  propensity score.
\newblock {\em Journal of the American Statistical Association} {\bf 99,}
  854--866.

\bibitem[\protect\citeauthoryear{Kennedy, Ma, McHugh, and Small}{Kennedy
  et~al.}{2017}]{kennedy2017non}
Kennedy, E.~H., Ma, Z., McHugh, M.~D., and Small, D.~S. (2017).
\newblock Non-parametric methods for doubly robust estimation of continuous
  treatment effects.
\newblock {\em Journal of the Royal Statistical Society: Series B (Statistical
  Methodology)} {\bf 79,} 1229--1245.

\bibitem[\protect\citeauthoryear{Kosorok}{Kosorok}{2008}]{kosorok2008introduction}
Kosorok, M.~R. (2008).
\newblock {\em Introduction to empirical processes and semiparametric
  inference}.
\newblock Springer Science \& Business Media.

\bibitem[\protect\citeauthoryear{Mu{\~n}oz and van~der Laan}{Mu{\~n}oz and
  van~der Laan}{2012}]{munoz2012population}
Mu{\~n}oz, I.~D. and van~der Laan, M. (2012).
\newblock Population intervention causal effects based on stochastic
  interventions.
\newblock {\em Biometrics} {\bf 68,} 541--549.

\bibitem[\protect\citeauthoryear{Neugebauer and van~der Laan}{Neugebauer and
  van~der Laan}{2007}]{neugebauer2007nonparametric}
Neugebauer, R. and van~der Laan, M. (2007).
\newblock Nonparametric causal effects based on marginal structural models.
\newblock {\em Journal of Statistical Planning and Inference} {\bf 137,}
  419--434.

\bibitem[\protect\citeauthoryear{Pfanzagl}{Pfanzagl}{1982}]{pfanzagl1982contributions}
Pfanzagl, J. (1982).
\newblock {\em Contributions to a general asymptotic statistical theory}.
\newblock Springer.

\bibitem[\protect\citeauthoryear{Robins, Hernan, and Brumback}{Robins
  et~al.}{2000}]{robins2000marginal}
Robins, J.~M., Hernan, M.~A., and Brumback, B. (2000).
\newblock Marginal structural models and causal inference in epidemiology.

\bibitem[\protect\citeauthoryear{Robins and Rotnitzky}{Robins and
  Rotnitzky}{1995}]{robins1995semiparametric}
Robins, J.~M. and Rotnitzky, A. (1995).
\newblock Semiparametric efficiency in multivariate regression models with
  missing data.
\newblock {\em Journal of the American Statistical Association} {\bf 90,}
  122--129.

\bibitem[\protect\citeauthoryear{Rubin}{Rubin}{1974}]{rubin1974estimating}
Rubin, D.~B. (1974).
\newblock Estimating causal effects of treatments in randomized and
  nonrandomized studies.
\newblock {\em Journal of educational Psychology} {\bf 66,} 688.

\bibitem[\protect\citeauthoryear{van~der Laan and Gruber}{van~der Laan and
  Gruber}{2016}]{van2016one}
van~der Laan, M. and Gruber, S. (2016).
\newblock One-step targeted minimum loss-based estimation based on universal
  least favorable one-dimensional submodels.
\newblock {\em The international journal of biostatistics} {\bf 12,} 351--378.

\bibitem[\protect\citeauthoryear{van~der Laan, Polley, and Hubbard}{van~der
  Laan et~al.}{2007}]{van2007super}
van~der Laan, M.~J., Polley, E.~C., and Hubbard, A.~E. (2007).
\newblock Super learner.
\newblock {\em Statistical applications in genetics and molecular biology} {\bf
  6,}.

\bibitem[\protect\citeauthoryear{van~der Laan and Rose}{van~der Laan and
  Rose}{2011}]{van2011targeted}
van~der Laan, M.~J. and Rose, S. (2011).
\newblock {\em Targeted learning: causal inference for observational and
  experimental data}.
\newblock Springer Science \& Business Media.

\bibitem[\protect\citeauthoryear{van~der Laan and Rose}{van~der Laan and
  Rose}{2018}]{van2018targeted}
van~der Laan, M.~J. and Rose, S. (2018).
\newblock {\em Targeted learning in data science}.
\newblock Springer.

\bibitem[\protect\citeauthoryear{van~der Laan and Rubin}{van~der Laan and
  Rubin}{2006}]{van2006targeted}
van~der Laan, M.~J. and Rubin, D. (2006).
\newblock Targeted maximum likelihood learning.
\newblock {\em The international journal of biostatistics} {\bf 2,}.

\bibitem[\protect\citeauthoryear{van~der Vaart and Wellner}{van~der Vaart and
  Wellner}{1996}]{van1996weak}
van~der Vaart, A. and Wellner, J. (1996).
\newblock {\em Weak convergence and empirical processes}.
\newblock Springer.

\bibitem[\protect\citeauthoryear{van~der Vaart}{van~der
  Vaart}{2000}]{van2000asymptotic}
van~der Vaart, A.~W. (2000).
\newblock {\em Asymptotic statistics}, volume~3.
\newblock Cambridge university press.

\bibitem[\protect\citeauthoryear{Wahba}{Wahba}{1990}]{wahba1990spline}
Wahba, G. (1990).
\newblock {\em Spline models for observational data}.
\newblock SIAM.

\bibitem[\protect\citeauthoryear{Westling}{Westling}{2021}]{westling2021nonparametric}
Westling, T. (2021).
\newblock Nonparametric tests of the causal null with nondiscrete exposures.
\newblock {\em Journal of the American Statistical Association} pages 1--12.

\bibitem[\protect\citeauthoryear{Westling, Gilbert, and Carone}{Westling
  et~al.}{2020}]{westling2020causal}
Westling, T., Gilbert, P., and Carone, M. (2020).
\newblock Causal isotonic regression.
\newblock {\em Journal of the Royal Statistical Society: Series B (Statistical
  Methodology)} {\bf 82,} 719--747.

\bibitem[\protect\citeauthoryear{Zhang, Zhou, Cao, and Zhang}{Zhang
  et~al.}{2016}]{zhang2016causal}
Zhang, Z., Zhou, J., Cao, W., and Zhang, J. (2016).
\newblock Causal inference with a quantitative exposure.
\newblock {\em Statistical methods in medical research} {\bf 25,} 315--335.

\bibitem[\protect\citeauthoryear{Zheng and van~der Laan}{Zheng and van~der
  Laan}{2011}]{zheng2011cross}
Zheng, W. and van~der Laan, M.~J. (2011).
\newblock Cross-validated targeted minimum-loss-based estimation.
\newblock In {\em Targeted Learning}, pages 459--474. Springer.

\end{thebibliography}

\newpage

\section*{Appendix}

\subsection*{Proofs of theoretical results}

\noindent \textbf{Proof of Lemma 1}
\\ 
Let $P$ be a distribution function that satisfies conditions A1 through A3, and let $p$ denote the denisty of $P$ with respect to a dominating measure $\mu$.
Let $\omega: \mathcal{W} \times \mathcal{A} \times \mathcal{Y} \to \mathbb{R}$ be an arbitrary function with zero mean and finite variance under $P$.
We define $P_\epsilon$ as the parametric submodel with density function $p_\epsilon: (w,a,y) \mapsto p(w,a,y)\{1 + \epsilon\omega(w,a,y)\}$ for $\epsilon$ small.
We can observe that $P_\epsilon$ is equal to $P$ at $\epsilon = 0$, and the score function is $\omega$.
Every regular parametric model that passes through $P$ and has score function equal to $\omega$ can be closely approximated using such a submodel.
If there exists function $\tilde{\phi}_P$ that satisfies
\begin{align}
\frac{d}{d\epsilon} \psi_{P_\epsilon, \theta^*}(h) \bigg \vert_{\epsilon = 0} = E_P[\tilde{\phi}_P(O;h) \omega(O)],
\label{eif-est-eqn}
\end{align}
for any choice of $\omega$, then $\psi_{P, \theta^*}(h)$ is pathwise differentiable in a nonparametric model, and $\tilde{\phi}_P(\cdot; h) - E_P[\tilde{\phi}_P(O; h)]$ is the nonparametric efficient influence function.
We show that such a function exists, and when centered about its mean, it is equal to $\phi_{P, \theta^*}(\cdot; h)$.

We first evaluate the derivative of $\psi_{P_\epsilon}$ at $\epsilon = 0$.
We express $\psi_{P_\epsilon, \theta^*}(h)$ as the sum of three components:
\begin{align*}
\psi_{P_\epsilon, \theta^*}(h) = G^{\mathrm{I}}_{P_\epsilon}(h) - G^{\mathrm{II}}_{P_\epsilon}(h) - G^{\mathrm{III}}_{P_\epsilon}(h), 
\end{align*}
where for any distribution $P$, we define 
\begin{align*}
&G^{\mathrm{I}}_P(h) := E_P[\theta_P(A) h(A)]
\\
&G^{\mathrm{II}}_P(h) := E_P[\theta_P(A)]E_P[h(A)]
\\
&G_P^{\mathrm{III}}(h) := E_P[\theta^*(A)h(A)] - E_P[\theta^*(A)]E_P[\theta^*(A)].
\end{align*}
To make calculation of the derivative of $\psi_{P_\epsilon, \theta^*}$ more manageable, we evaluate the derivative of the three additive components at $\epsilon = 0$.

We first calculate $\frac{d}{d\epsilon} G^{\mathrm{I}}_{P_\epsilon}(h) \vert|_{\epsilon = 0}$.
We can write $G^{\mathrm{I}}_{P_\epsilon}(h)$ as
\begin{align*}
G^{\mathrm{I}}_{P_\epsilon}(h) = \int \theta_{P_\epsilon}(a) h(a) p(w,a,y)\{1 + \epsilon\omega(w,a,y)\}\} d\mu(w,a,y),
\end{align*}
where $\theta_{P_\epsilon}$ is the dose-response function under $P_\epsilon$.
To proceed, we need an expression for the derivative of $\theta_{P_\epsilon}(a)$ with respect to $\epsilon$.
We can write 
\begin{align*}
\theta_{P_\epsilon}(a) = \left[\int Q_{P_{\epsilon}}(w,a)\left\{\int p(w,a,y)\left\{1 + \epsilon\omega(w,a,y)\right\} \mu(dy, da)\right\}\mu(dw) \right],
\end{align*}
where $Q_{P_\epsilon}$ is the conditional mean of $Y$ given $A$ and $W$ under $P_\epsilon$.
Letting $p^\mathrm{i}(y|a,w)$ be the conditional density of $Y$ given $A$ and $W$, it can be shown that the derivative of $Q_{P_\epsilon}$ at $\epsilon = 0$ is
\begin{align*}
\frac{d}{d\epsilon} Q_{P_\epsilon}(w,a)\bigg|_{\epsilon = 0}  =&\int \left\{y - Q_P(w,a)\right\}\omega(w,a,y)p^\mathrm{i}(y|a,w)d\mu(y) .
\end{align*}
Now, let the conditional density of $(A,Y)$ given $W$ be $p^{\mathrm{ii}}(a,y|w)$, and let the marginal density of $W$ be $p^{\mathrm{iii}}(w)$.
We now calculate the derivative of $\theta_{P_\epsilon}(a)$ at $\epsilon = 0$:
\begin{align}
\frac{d}{d\epsilon} \theta_{P_\epsilon}(a) \bigg|_{\epsilon = 0} = &\int  \left[\frac{d}{d\epsilon} Q_{P_\epsilon}(w,a) \right]_{\epsilon = 0} \left\{\int p(w,a,y)\mu(dy,da)\right\} \mu(dw) + \nonumber
\\
&\int Q_P(w,a) \left\{\int p(w,a,y) \omega(w,a,y)\mu(dy, da)\right\}\mu(dw) \nonumber
\\ 
\nonumber
\\
= &\int  \left[\int \left\{y - Q_P(w,a)\right\}\omega(w,a,y)p^\mathrm{i}(y|a,w)d\mu(y) \right] p^{\mathrm{iii}}(w) \mu(dw) + \nonumber
\\
&\int Q_P(w,a)p^{\mathrm{iii}}(w) \left\{\int p^{\mathrm{ii}}(a,y|w) \omega(w,a,y)\mu(dy, da)\right\}\mu(dw) .
\label{deriv-drf}
\end{align}
We can now express the derivative of $G^{\mathrm{I}}_{P_\epsilon}(h)$ as
\begin{align*}
\frac{d}{d\epsilon} G^{\mathrm{I}}_{P_\epsilon}(h) \bigg \vert_{\epsilon = 0} = &\int \frac{d}{d\epsilon}\bigg[\theta_{P_\epsilon}(a)\bigg]_{\epsilon = 0} h(a) p(w,a,y)d\mu(w,a,y)  + \int \theta_P(a) p(w,a,y)\omega(w,a,y)d\mu(w,a,y) 
\\
\\
= &\int\int  \left[\int \left\{y - Q_P(w,a)\right\}\omega(w,a,y)p^\mathrm{i}(y|a,w)d\mu(y) \right] p^{\mathrm{iii}}(w) \mu(dw)  h(a) p(w,a,y)d\mu(w,a,y)   +
\\
&\int\int \int Q_P(w,a)\left\{\int q_0^{\mathrm{ii}}(a,y|w) \omega(w,a,y)\mu(dy, da)\right\}p^{\mathrm{iii}}(w) \mu(dw)  h(a)p(w,a,y)d\mu(w,a,y) +
\\
&\int \theta_P(a)h(a) p(w,a,y)\omega(w,a,y)d\mu(w,a,y).
\end{align*}
Let the marginal density of $A$ be $p^{\mathrm{iv}}(a)$, and let the conditional density of $W$ given $A$ be $p^{\mathrm{v}}(w|a)$.
Observe that we can express the marginal density of $W$ as
\begin{align*}
p^{\mathrm{iii}}(w) = p^{\mathrm{v}}(w|a)\left\{\frac{p^{\mathrm{iv}}(a)}{g_P(a|w)} \right\}.
\end{align*}
Using this fact, we can write
\begin{align}
\frac{d}{d\epsilon} G^{\mathrm{I}}_{P_\epsilon}(h) \bigg\vert_{\epsilon = 0} = &E_P\Bigg[ \bigg[ \left\{ \frac{p^{\textrm{iv}}(A)}{g_P(A|W)}\right\} E_0[\left\{Y - Q_P(W,A)\right\}\omega(W,A,Y)|W,A]  \bigg| A \bigg] h(A) \Bigg]  \nonumber
+ 
\\
&E_P\Bigg[ \bigg[\left\{ \frac{p^{\textrm{iv}}(A)}{g_P(A|W)}\right\} Q_P(W,A)E_P[\omega(W,A,Y)|W]  \bigg| A \bigg] h(A) \Bigg] + \nonumber
\\
&E_P[\theta_P(A) h(A) \omega(W,A,Y)].
\label{eif-proof-1}
\end{align}
By applying the law of total expectation in the first two lines, we get
\begin{align*}
\frac{d}{d\epsilon} G^{\mathrm{I}}_{P_\epsilon}(h) \bigg\vert_{\epsilon = 0} = &E_P\Bigg[  \left\{ \frac{p^{\textrm{iv}}(A)}{g_P(A|W)}\right\} \left\{Y - Q_P(W,A)\right\}  h(A)\omega(W,A,Y) \Bigg] 
+ 
\\
&E_P\Bigg[ E_P\bigg[\left\{ \frac{p^{\textrm{iv}}(A)}{g_P(A|W)}\right\} Q_P(W,A)h(A)\bigg|W\bigg]    \omega(W,A,Y) \Bigg] +
\\
&E_P[\theta_P(A) h(A) \omega(W,A,Y)].
\end{align*}
Finally, by observing that in the second line above, the conditional densities of $A$ given $W$ cancel, we have
\begin{align*}
\frac{d}{d\epsilon} G^{\mathrm{I}}_{P_\epsilon}(h) \bigg\vert_{\epsilon = 0} = &E_P\Bigg[  \left\{ \frac{p^{\textrm{iv}}(A)}{g_P(A|W)}\right\} \left\{Y - Q_P(W,A)\right\}  h(A)\omega(W,A,Y) \Bigg] 
+ 
\\
&E_P\bigg[ E_{P,A}\left[ Q_P(W,A)h(A)\right]    \omega(W,A,Y) \bigg] +
\\
&E_P[\theta_P(A) h(A) \omega(W,A,Y)],
\end{align*} 
where $E_{P,A}$ is the expectation over the marginal distribution of $A$ under $P$.

We now evaluate the derivative of $G^{\mathrm{II}}_{P_\epsilon}(h)$.
We can express $G^{\mathrm{II}}_{P_\epsilon}$ as
\begin{align*}
G^{\mathrm{II}}_{P_\epsilon}(h) = 
&\left[\int \theta_{P_\epsilon}(a)p(w,a,y)\{1 + \epsilon\omega(w,a,y)\} d\mu(w,a,y)\right] \times
\\
&\left[
\int h(a) p(w,a,y)\left\{1 + \epsilon \omega(w,a,y)\right\}d\mu(w,a,y)
\right].
\end{align*}
The evaluation of its derivative at zero is
\begin{align*}
\frac{d}{d\epsilon} G^{\mathrm{II}}_{P_\epsilon}(h) \bigg|_{\epsilon = 0} 
=
&
\left\{\int h(a) p(w,a,y) d\mu(w,a,y)\right\}\left\{\int \frac{d}{d\epsilon}\big[\theta_{P_\epsilon}(a) \big]_{\epsilon = 0}p(w,a,y)d\mu(w,a,y)\right\}+ 
\\
&\left\{\int h(a) p(w,a,y) d\mu(w,a,y)\right\}\left\{\int \theta_P(a) p(w,a,y) \omega(w,a,y) d\mu(w,a,y)\right\} +
\\
&\left\{\int h(a) \omega(w,a,y) p(w,a,y) d\mu(w,a,y)\right\} \left\{\int \theta_P(a) p(w,a,y)d\mu(w,a,y)\right\}.
\end{align*}
Performing similar steps as were used to calculate $\frac{d}{d\epsilon}G^{\mathrm{I}}_{P_\epsilon}(h) \bigg |_{\epsilon = 0}$, it can be shown that
\begin{align}
\frac{d}{d\epsilon}G^{\mathrm{II}}_{P_\epsilon}(h) \bigg|_{\epsilon = 0} =
&E_P[h(A)]E_0\left[\frac{p^{\mathrm{iv}}(A)}{g_P(A|W)}\left\{Y - Q_P(W,A)\right\}\omega(W,A,Y)\right] + \nonumber
\\
&E_P\bigg[E_P[h(A)]E_{P,A}[Q_P(W,A)] \omega(W,A,Y)\bigg] + \nonumber
\\
 &E_P\bigg[E_P[h(A)] \theta_P(A)\omega(W,A,Y)\bigg] + E_P\bigg[E_P[\theta_P(A)] h(A)\omega(W,A,Y)\bigg].
 \label{eif-proof-2}
\end{align}

Now, we take the derivative of the remaining term $G^{\mathrm{III}}_{P_\epsilon}(h)$.
Rather than perform this calculation  directly, we recognize that $G^{\mathrm{III}}_P(h)$ is simply the covariance between $h(A)$ and $\theta^*(A)$ under $P$, and it is well known that the derivative can be expressed as
\begin{align}
\frac{d}{d\epsilon} G^{\mathrm{III}}_{P_\epsilon}(h) \bigg|_{\epsilon = 0} &= E_P\bigg[\left\{\theta^*(A)h(A) - E[h(A)]\theta^*(A) - E[\theta^*(A)]h(A) \right\}\omega(W,A,Y)\bigg].
\label{eif-proof-3}
\end{align}


Now, by \eqref{eif-proof-1}, \eqref{eif-proof-2}, and \eqref{eif-proof-3}, we can express $\frac{d}{d\epsilon}\psi_{P_\epsilon, \theta^*}(h) \big\vert_{\epsilon = 0}$ as
\begin{align*}
\frac{d}{d\epsilon}\psi_{P_\epsilon, \theta^*}(h) \big\vert_{\epsilon = 0} &= \frac{d}{d\epsilon} \bigg[G^{\mathrm{I}}_{P_\epsilon}(h) - G^{\mathrm{II}}_{P_\epsilon}(h)  - G^{\mathrm{III}}_{P_\epsilon}(h) \bigg]_{\epsilon = 0}
\\
&=
E_P\left[\phi_P^*(W,A,Y;h)\omega(W,A,Y)\right],
\end{align*}
where we define
\begin{align*}
\phi_P^*(w,a,y; h) := &\left[\bar{\theta}_P(a) - \bar{\theta}^*_P(a)  + \frac{p^{\textit{iv}}(a)}{g_P(a|w)} \left\{y - Q_P(w,a)   \right\} \right]\left\{h(a) - E_P[h(A)]\right\}
 +
\\
&E_P\left[Q_P(w,A)\left\{h(A) - E_P[h(A)]\right\}\right].
\end{align*}
The proof is completed by observing that $p^{\textrm{iv}}(a) = E_P[g_{P}(a|W)]$, 
and
\begin{align*}
\phi_{P,\theta^*}(\cdot;h) = \phi_P^*(\cdot; h) - E_P[\phi_P^*(W,A,Y; h)].
\end{align*}

\noindent \textbf{Proof of Lemma 2.}

This result is an immediate consequence of Slutsky's theorem  (see, e.g., Theorem 7.15 of \citealp{kosorok2008introduction}).

\noindent \textbf{Proof of Theorem 1.}


The one-step estimator can be expressed as
\begin{align*}
\psi^{\mathrm{I}}_{n, \theta^*}(h) - \psi_{0, \theta^*}(h)  = n^{-1} \sum_{i=1}^n \phi_{P_0}(O_i; h) + R_{1,n}(h) + R_{2,n}(h),
\end{align*} 
where we define
\begin{align*}
R_{1,n}(h) &:= -\left[n^{-1} \sum_{i=1}^n \phi_{\hat{P},\theta^*}(O_i; h) - \phi_{P_0, \theta^*}(O_i; h) - \left\{\int \phi_{\hat{P}}(o; h) - \phi_{P_0}(o; h) dP_0(o)\right\}\right],
\\
R_{2,n}(h) &:= \psi_{n, \theta^*}(h) - \psi_{0, \theta^*}(h)  -  \int \phi_{\hat{P}, \theta^*}(o; h) dP_0(o).
\end{align*}
By the triangle inequality, it is sufficient to show that $\sup_{h \in \mathcal{H}}|R_{1,n}(h)|$ and $\sup_{h \in \mathcal{H}}|R_{2,n}(h)|$ are both $o_P(n^{-1/2})$.

To see that $\sup|R_{1,n}(h)| = o_P(n^{-1/2})$, we note that by Assumption B1 and Theorem 2.10.6  of \cite{van1996weak}, $\mathcal{H}$ is a $P_0$-Donsker class, and therefore,
 $\sup_{h \in \mathcal{H}} \left|n^{-1}\sum_{i=1}^n h(A_i) - E_0[h(A)]\right| = O_P(n^{-1/2})$. \sloppy
In view of this fact and Assumption B2, it can be concluded that the estimator $\phi_{\hat{P}_n, \theta^*}$ of the efficient influence function is uniformly consistent over $\mathcal{H}$, that is
\begin{align*}
\sup_{h \in \mathcal{H}}\int \left\{\phi_{\hat{P}_n}(o;h) -  \phi_{0, \theta^*}(o; h)\right\}^2 dP_0(o) = o_P(1). 
\end{align*}
It is shown in the proof of lemma 19.26 in \cite{van2000asymptotic} that when this condition is satisfied, and Assumption B2 holds,  $\sup_{h \in \mathcal{H}} \left| R_{1,n}(h) \right| = o_P(n^{-1/2})$.
%


It remains to be shown that $R_{2,n}(h)$ is asymptotically negligible.
With some algebra, it can be shown that the second remainder term can be expressed as
\begin{align*}
R_{2,n}(h) = R_{2,n}^{\mathrm{i}}(h) + R_{2,n}^{\mathrm{ii}}(h)  + R_{2,n}^{\mathrm{iii}}(h)  + R_{2,n}^{\mathrm{iv}}(h)  + R_{2,n}^{\mathrm{v}}(h),
\end{align*}
where we define
\begin{align*}
R^{\mathrm{i}}_{2,n}(h) :=  &\int \left\{\frac{E_0[g_0(a|W)]}{g_0(a|w)}  - \frac{n^{-1}\sum_{i=1}^ng_n(a|W_i)}{g_n(a|w)} \right\} 
 \left\{ Q_{0}(w,a) - Q_{n}(w,a) \right\} \left\{h(a) - E_0[h(A)]\right\} dP_0(w,a)
  \\
R^{\mathrm{ii}}_{2,n}(h) := & n^{-1} \sum_{i=1}^n \left\{ n^{-1}\sum_{j=1}^n Q_n(W_i, A_j)\left\{h(A_j) - E_0[h(A)] \right\} - \int Q_n(W_i, a)\left\{h(a) - E_0[h(A)] \right\} dP_0(a)\right\} - 
\\
&\int \left\{ n^{-1}\sum_{j=1}^n Q_n(w, A_j)\left\{h(A_j) - E_0[h(A_j)]  \right\} - \int Q_n(w, a)\left\{h(a) - E_0[h(A)]\right\} dP_0(a)\right\}dP_0(w)
\\
R^{\mathrm{iii}}_{2,n}(h) := &\left\{ n^{-1}\sum_{i=1}^n h(A_i) - E_0[h(A)] \right\}  \int \frac{n^{-1}\sum_{i=1}^ng_n(a|W_i)}{g_n(a|w)}\left\{Q_0(w,a) - Q_n(w,a)\right\}dP_0(a)
\\
R^{\mathrm{iv}}_{2,n}(h) := &\left\{E_0[h(A)] - n^{-1}\sum_{i=1}^nh(A_i) \right\} 
\left[ n^{-1} \sum_{i=1}^n \left\{ n^{-1}\sum_{j=1}^n Q_n(W_i, A_j) - \int Q_n(W_i, a) dP_0(a)\right\}\right] -
 \\
&\left\{E_0[h(A)] - n^{-1}\sum_{i=1}^n h(A_i) \right\} 
\left[
\int \left\{ n^{-1}\sum_{j=1}^n Q_n(w, A_j) - \int Q_n(w, a) dP_0(a)\right\}dP_0(w)
\right]
\\
R^{\mathrm{v}}_{2,n}(h) := &\left\{n^{-1}\sum_{i=1}^n h(A_i) - E_0[h(A)]\right\}\left\{n^{-1}\sum_{i=1}^n \theta^*(A_i) - E_0[\theta^*(A)]\right\}.
\end{align*}
By the triangle inequality, it suffices to argue that each of the above components converges to zero in probability at a rate of $n^{1/2}$ uniformly in $\mathcal{H}$.

It follows directly from Assumption B1 that $\sup_{h \in \mathcal{H}}|R_{2,n}^{\mathrm{i}}(h)| = o_P(n^{-1/2})$.
We now argue that $\sup_{h \in \mathcal{H}}\left|R_{2,n}^{\mathrm{ii}}(h)\right| = o_P(n^{-1/2})$.
Observe that 
\begin{align*}
&\sup_{h \in \mathcal{H}}\int  \left\{n^{-1}\sum_{i=1}^n Q_{n}(w, A_i)h(A_i) - \int Q_n(w,a) h(a)dP_0(a) \right\}^2 dP_0(w) \leq 
\\
&2\sup_{h \in \mathcal{H}}\int  \left[n^{-1}\sum_{i=1}^n \left\{Q_{n}(w, A_i) - Q_0(w, A_i) \right\}h(A_i) - \int \left\{Q_n(w,a) - Q_0(w,a) \right\}h(a) dP_0(a) \right]^2 dP_0(w) + 
\\
&2\sup_{h \in \mathcal{H}}\int  \left\{n^{-1}\sum_{i=1}^n Q_{0}(w, A_i)h(A_i) - \int Q_0(w,a)h(a) dP_0(a) \right\}^2 dP_0(w) = o_P(1),
\end{align*}
where the convergence follows from the rate conditions and the Donsker assumption.
It can thus be concluded using the argument in the proof of Lemma 19.26 of \cite{van2000asymptotic} that $\sup_{h \in \mathcal{H}}\left|R^{\mathrm{ii}}_{2,n}(h) \right| = o_P(n^{-1/2})$.
It can be seen that $\sup_{h \in \mathcal{H}}|R^{\mathrm{iii}}_n(h)| = o_P(n^{-1/2})$ and $\sup_{h \in \mathcal{H}}|R^{\mathrm{iv}}_n(h)| = o_P(n^{-1/2})$ by recalling that $\sup_{h \in \mathcal{H}}|n^{-1}\sum_{i=1}^n h(A_i) - E_0[h(A)]| = O_P(n^{-1/2})$ and noting that $Q_{n}$ is a consistent estimator for $Q_0$.
Similarly, that $\sup_{h \in \mathcal{H}}|R^{\mathrm{v}}_n(h)| = o_P(n^{-1/2})$ follows from uniform consistency of $n^{-1}\sum_{i=1}^n h(A_i)$.
This completes our argument to show that the one-step estimator is uniformly asymptotically linear.

\noindent \textbf{Proof of Theorem 2.}

The proof of Theorem 2 is nearly the same as the proof of Theorem 1, so we only provide a brief outline of the argument. 
The TML estimator can be expressed as
\begin{align*}
\psi^{\mathrm{II}}_{n, \theta^*}(h) - \psi_{0, \theta^*}(h)  = n^{-1} \sum_{i=1}^n \phi_{P_0}(O_i; h) + R_{1,n}(h) + R_{2,n}(h) + R_{3,n}(h),
\end{align*} 
where we define
\begin{align*}
R_{1,n}(h) &:= -\left[n^{-1} \sum_{i=1}^n \phi_{\tilde{P},\theta^*}(O_i; h) - \phi_{P_0, \theta^*}(O_i; h) - \left\{\int \phi_{\tilde{P}}(o; h) - \phi_{P_0}(o; h) dP_0(o)\right\}\right],
\\
R_{2,n}(h) &:= \psi_{n, \theta^*}(h) - \psi_{0, \theta^*}(h)  -  \int \phi_{\tilde{P}, \theta^*}(o; h) dP_0(o)
\\
R_{3,n}(h) &:= -n^{-1} \sum_{i=1}^n \phi_{\tilde{P}_n}(O_i; h).
\end{align*}
Uniform asymptotic negligibility of $R_{1,n}(h)$ and $R_{2,n}(h)$ follows from the same arguments we presented for establishing uniform asymptotic linearity of the one-step estimator.
Additionally, we have by construction that $\sup_{h \in \mathcal{H}} |R_{3,n}(h)|  = o_P(n^{-1/2})$.
From this, we can conclude that the TML estimator is uniformly asymptotically linear.

\end{document}